\long\def\del#1\enddel{}
\long\def\new#1\endnew{{\bf #1}}	\long\def\new#1\endnew{{#1}}

\font\sixrm=cmr6 at 6pt
\font\eightrm=cmr8 at 8pt

\font\seventeenrm=cmr17 at 17pt
\font\twentyonerm=cmr17 at 21pt

\font\ssten=cmss10

\font\twelvecal=cmsy10 at 12pt

\font\twelvemath=cmmi12

\font\seventeenbold=cmbx7 at 17pt

\font\fively=lasy5
\font\sevenly=lasy7
\font\tenly=lasy10

\textfont10=\tenly
\scriptfont10=\sevenly
\scriptscriptfont10=\fively
\magnification=1180
\parskip=10pt
\parindent=20pt
\def\today{\ifcase\month\or January\or February\or March\or April\or May\or
June
       \or July\or August\or September\or October\or November\or December\fi
       \space\number\day, \number\year}

\def\title#1{\footline={\ifnum\pageno<2\hfil
       \else\hss\tenrm\folio\hss\fi}\vskip1truein\centerline{{#1}
       \footnote{\raise1ex\hbox{*}}{\eightrm Supported in part
       by the Robert A. Welch Foundation and N.S.F. Grants
       PHY-880637 and\break PHY-8605978.}}}

\def\newpage{\vfill\eject}
\def\abstract#1{\centerline{\bf ABSTRACT}\vskip.2truein{\narrower\noindent#1
       \smallskip}}

\def\runninghead#1#2{\voffset=2\baselineskip\nopagenumbers
       \headline={\ifodd\pageno\rightheadline\else \leftheadline\fi}
       \def\rightheadline{{\sl#1}\hfill{\rm\folio}}
       \def\leftheadline{{\rm\folio}\hfill{\sl#2}}}
\def\SS{\mathhexbox278}

\newcount\footnoteno
\def\Footnote#1{\advance\footnoteno by 1
                \let\SF=\empty
                \ifhmode\edef\SF{\spacefactor=\the\spacefactor}\/\fi
                $^{\the\footnoteno}$\ignorespaces
                \SF\vfootnote{$^{\the\footnoteno}$}{#1}}

\def\place#1#2#3{\vbox to0pt{\kern-\parskip\kern-7pt
                             \kern-#2truein\hbox{\kern#1truein #3}
                             \vss}\nointerlineskip}
\def\figurecaption#1#2{\kern.75truein\vbox{\hsize=5truein\noindent{\bf Figure
    \figlabel{#1}:} #2}}
\def\tablecaption#1#2{\kern.75truein\lower12truept\hbox{\vbox{\hsize=5truein
    \noindent{\bf Table\hskip5truept\tablabel{#1}:} #2}}}
\def\boxed#1{\lower3pt\hbox{
                       \vbox{\hrule\hbox{\vrule

\vbox{\kern2pt\hbox{\kern3pt#1\kern3pt}\kern3pt}\vrule}
                         \hrule}}}

\def\g{\gamma}\def\G{\Gamma}
\def\d{\delta}\def\D{\Delta}

\def\l{\lambda}

\def\S{\Sigma}

\def\ca#1{\relax\ifmmode {{\cal #1}}\else $\cal #1$\fi}

\def\calb{{\cal B}}

\def\calm{{\cal M}}

\def\inbar{\vrule height1.5ex width.4pt depth0pt}
\def\IB{\relax{\rm I\kern-.18em B}}
\def\IC{\relax\hbox{\kern.25em$\inbar\kern-.3em{\rm C}$}}
\def\ID{\relax{\rm I\kern-.18em D}}
\def\IE{\relax{\rm I\kern-.18em E}}
\def\IF{\relax{\rm I\kern-.18em F}}
\def\IG{\relax\hbox{\kern.25em$\inbar\kern-.3em{\rm G}$}}
\def\IH{\relax{\rm I\kern-.18em H}}
\def\II{\relax{\rm I\kern-.18em I}}
\def\IK{\relax{\rm I\kern-.18em K}}
\def\IL{\relax{\rm I\kern-.18em L}}
\def\IM{\relax{\rm I\kern-.18em M}}
\def\IN{\relax{\rm I\kern-.18em N}}
\def\IO{\relax\hbox{\kern.25em$\inbar\kern-.3em{\rm O}$}}
\def\IP{\relax{\rm I\kern-.18em P}}
\def\IQ{\relax\hbox{\kern.25em$\inbar\kern-.3em{\rm Q}$}}
\def\IR{\relax{\rm I\kern-.18em R}}
\def\IZ{\relax\ifmmode\hbox{\ssten Z\kern-.4em Z}\else{\ssten Z\kern-.4em Z}
\fi}
\def\IGa{\relax{\rm I}\kern-.18em\Gamma}
\def\IPi{\relax{\rm I}\kern-.18em\Pi}
\def\ITh{\relax\hbox{\kern.25em$\inbar\kern-.3em\Theta$}}
\def\IOm{\relax\thinspace\inbar\kern1.95pt\inbar\kern-5.525pt\Omega}


\def\cy{Calabi--Yau}
\def\cym{Calabi--Yau manifold}
\def\cys{Calabi--Yau manifolds}

\def\K{K\"ahler}

\def\H#1#2{\relax\ifmmode {H^{#1#2}}\else $H^{#1 #2}$\fi}
\def\M{\relax\ifmmode{\calm}\else $\calm$\fi}

\def\Bigcheck{\lower3.8pt\hbox{\smash{\hbox{{\twentyonerm \v{}}}}}}
\def\bigboldcheck{\smash{\hbox{{\seventeenbold\v{}}}}}

\def\Bighat{\lower3.8pt\hbox{\smash{\hbox{{\twentyonerm \^{}}}}}}

\def\Msharp{\relax\ifmmode{\calm^\sharp}\else $\smash{\calm^\sharp}$\fi}
\def\Mflat{\relax\ifmmode{\calm^\flat}\else $\smash{\calm^\flat}$\fi}
\def\preMcheck{\kern2pt\hbox{\Bigcheck\kern-12pt{$\cal M$}}}
\def\Mcheck{\relax\ifmmode\preMcheck\else $\preMcheck$\fi}
\def\preMhat{\kern2pt\hbox{\Bighat\kern-12pt{$\cal M$}}}
\def\Mhat{\relax\ifmmode\preMhat\else $\preMhat$\fi}

\def\Bsharp{\relax\ifmmode{\calb^\sharp}\else $\calb^\sharp$\fi}
\def\Bflat{\relax\ifmmode{\calb^\flat}\else $\calb^\flat$ \fi}
\def\preBcheck{\hbox{\Bigcheck\kern-9pt{$\cal B$}}}
\def\Bcheck{\relax\ifmmode\preBcheck\else $\preBcheck$\fi}
\def\preBhat{\hbox{\Bighat\kern-9pt{$\cal B$}}}
\def\Bhat{\relax\ifmmode\preBhat\else $\preBhat$\fi}

\def\figBcheck{\kern3pt\hbox{\raise1pt\hbox{\bigboldcheck}\kern-11pt
    {\twelvecal B}}}
\def\figBsharp{{\twelvecal B}\raise5pt\hbox{$\twelvemath\sharp$}}
\def\figBflat{{\twelvecal B}\raise5pt\hbox{$\twelvemath\flat$}}

\def\gcheck{\hbox{\lower2.5pt\hbox{\Bigcheck}\kern-8pt$\g$}}
\def\lhat{\hbox{\raise.5pt\hbox{\Bighat}\kern-8pt$\l$}}

\def\Fcheck{\kern2pt\hbox{\raise1pt\hbox{\Bigcheck}\kern-10pt{$\cal F$}}}
\def\Fhat{\kern2pt\hbox{\raise1pt\hbox{\Bighat}\kern-10pt{$\cal F$}}}

\def\cp#1{\relax\ifmmode {\IP\kern-2pt{}_{#1}}\else $\IP\kern-2pt{}_{#1}$\fi}
\def\h#1#2{\relax\ifmmode {b_{#1#2}}\else $b_{#1#2}$\fi}

\def\frac#1#2{{#1\over #2}}

\def\cone{\relax\thinspace\hbox{$<\kern-.8em{)}$}}
\mathchardef\mho"0A30

\def\-{\hphantom{-}}

\def\ip{\amalg}


\def\npb#1{Nucl.\ Phys.\ {\bf B#1}}

\def\plb#1{Phys. Lett. {\bf #1B}}


\newif\ifproofmode
\proofmodefalse

\newif\ifforwardreference
\forwardreferencefalse

\newif\ifchapternumbers
\chapternumbersfalse

\newif\ifcontinuousnumbering
\continuousnumberingfalse

\newif\iffigurechapternumbers
\figurechapternumbersfalse

\newif\ifcontinuousfigurenumbering
\continuousfigurenumberingfalse

\newif\iftablechapternumbers
\tablechapternumbersfalse

\newif\ifcontinuoustablenumbering
\continuoustablenumberingfalse

\font\eqsixrm=cmr6

\font\sixrm=cmr6 at 6pt

\def\marginstyle{\eqsixrm}

\newtoks\chapletter
\newcount\chapno
\newcount\eqlabelno
\newcount\figureno
\newcount\tableno

\chapno=0
\eqlabelno=0
\figureno=0
\tableno=0

\def\chapfolio{\ifnum\chapno>0 \the\chapno\else\the\chapletter\fi}

\def\bumpchapno{\ifnum\chapno>-1 \global\advance\chapno by 1
\else\global\advance\chapno by -1 \setletter\chapno\fi
\ifcontinuousnumbering\else\global\eqlabelno=0 \fi
\ifcontinuousfigurenumbering\else\global\figureno=0 \fi
\ifcontinuoustablenumbering\else\global\tableno=0 \fi}

\def\setletter#1{\ifcase-#1{}\or{}%
\or\global\chapletter={A}%
\or\global\chapletter={B}%
\or\global\chapletter={C}%
\or\global\chapletter={D}%
\or\global\chapletter={E}%
\or\global\chapletter={F}%
\or\global\chapletter={G}%
\or\global\chapletter={H}%
\or\global\chapletter={I}%
\or\global\chapletter={J}%
\or\global\chapletter={K}%
\or\global\chapletter={L}%
\or\global\chapletter={M}%
\or\global\chapletter={N}%
\or\global\chapletter={O}%
\or\global\chapletter={P}%
\or\global\chapletter={Q}%
\or\global\chapletter={R}%
\or\global\chapletter={S}%
\or\global\chapletter={T}%
\or\global\chapletter={U}%
\or\global\chapletter={V}%
\or\global\chapletter={W}%
\or\global\chapletter={X}%
\or\global\chapletter={Y}%
\or\global\chapletter={Z}\fi}

\def\tempsetletter#1{\ifcase-#1{}\or{}%
\or\global\chapletter={A}%
\or\global\chapletter={B}%
\or\global\chapletter={C}%
\or\global\chapletter={D}%
\or\global\chapletter={E}%
\or\global\chapletter={F}%
\or\global\chapletter={G}%
\or\global\chapletter={H}%
\or\global\chapletter={I}%
\or\global\chapletter={J}%
\or\global\chapletter={K}%
\or\global\chapletter={L}%
\or\global\chapletter={M}%
\or\global\chapletter={N}%
\or\global\chapletter={O}%
\or\global\chapletter={P}%
\or\global\chapletter={Q}%
\or\global\chapletter={R}%
\or\global\chapletter={S}%
\or\global\chapletter={T}%
\or\global\chapletter={U}%
\or\global\chapletter={V}%
\or\global\chapletter={W}%
\or\global\chapletter={X}%
\or\global\chapletter={Y}%
\or\global\chapletter={Z}\fi}

\def\chapshow#1{\ifnum#1>0 \relax#1%
\else{\tempsetletter{\number#1}\chapno=#1\chapfolio}\fi}

\def\chaplabel#1{\bumpchapno\ifproofmode\ifforwardreference
\immediate\write1{\noexpand\expandafter\noexpand\def
\noexpand\csname CHAPLABEL#1\endcsname{\the\chapno}}\fi\fi
\global\expandafter\edef\csname CHAPLABEL#1\endcsname
{\the\chapno}\ifproofmode\llap{\hbox{\marginstyle #1\ }}\fi\chapfolio}

\def\eqnum{\global\advance\eqlabelno by 1
\eqno(\ifchapternumbers\chapfolio.\fi\the\eqlabelno)}

\def\eqlabel#1{\global\advance\eqlabelno by 1 \ifproofmode\ifforwardreference
\immediate\write1{\noexpand\expandafter\noexpand\def
\noexpand\csname EQLABEL#1\endcsname{\the\chapno.\the\eqlabelno?}}\fi\fi
\global\expandafter\edef\csname EQLABEL#1\endcsname
{\the\chapno.\the\eqlabelno?}\eqno(\ifchapternumbers\chapfolio.\fi
\the\eqlabelno)\ifproofmode\rlap{\hbox{\marginstyle #1}}\fi}

\def\eqalignnum{\global\advance\eqlabelno by 1
&(\ifchapternumbers\chapfolio.\fi\the\eqlabelno)}

\def\eqalignlabel#1{\global\advance\eqlabelno by 1 \ifproofmode
\ifforwardreference\immediate\write1{\noexpand\expandafter\noexpand\def
\noexpand\csname EQLABEL#1\endcsname{\the\chapno.\the\eqlabelno?}}\fi\fi
\global\expandafter\edef\csname EQLABEL#1\endcsname
{\the\chapno.\the\eqlabelno?}&(\ifchapternumbers\chapfolio.\fi
\the\eqlabelno)\ifproofmode\rlap{\hbox{\marginstyle #1}}\fi}

\def\eqref#1{\hbox{(\ifundefined{EQLABEL#1}***)\ifproofmode\ifforwardreference%
\else\write16{ ***Undefined Equation Reference #1*** }\fi
\else\write16{ ***Undefined Equation Reference #1*** }\fi
\else\edef\LABxx{\getlabel{EQLABEL#1}}%
\def\LAByy{\expandafter\stripchap\LABxx}\ifchapternumbers%
\chapshow{\LAByy}.\expandafter\stripeq\LABxx%
\else\ifnum\number\LAByy=\chapno\relax\expandafter\stripeq\LABxx%
\else\chapshow{\LAByy}.\expandafter\stripeq\LABxx\fi\fi)\fi}%
\ifproofmode\write2{Equation #1}\fi}

\def\fignum{\global\advance\figureno by 1
\relax\iffigurechapternumbers\chapfolio.\fi\the\figureno}

\def\figlabel#1{\global\advance\figureno by 1
\relax\ifproofmode\ifforwardreference
\immediate\write1{\noexpand\expandafter\noexpand\def
\noexpand\csname FIGLABEL#1\endcsname{\the\chapno.\the\figureno?}}\fi\fi
\global\expandafter\edef\csname FIGLABEL#1\endcsname
{\the\chapno.\the\figureno?}\iffigurechapternumbers\chapfolio.\fi
\ifproofmode\llap{\hbox{\marginstyle#1
\kern1.2truein}}\relax\fi\the\figureno}

\def\figref#1{\hbox{\ifundefined{FIGLABEL#1}!!!!\ifproofmode%
\ifforwardreference%
\else\write16{ ***Undefined Figure Reference #1*** }\fi
\else\write16{ ***Undefined Figure Reference #1*** }\fi
\else\edef\LABxx{\getlabel{FIGLABEL#1}}%
\def\LAByy{\expandafter\stripchap\LABxx}\iffigurechapternumbers%
\chapshow{\LAByy}.\expandafter\stripeq\LABxx%
\else\ifnum \number\LAByy=\chapno\relax\expandafter\stripeq\LABxx%
\else\chapshow{\LAByy}.\expandafter\stripeq\LABxx\fi\fi\fi}%
\ifproofmode\write2{Figure #1}\fi}

\def\tabnum{\global\advance\tableno by 1
\relax\iftablechapternumbers\chapfolio.\fi\the\tableno}

\def\tablabel#1{\global\advance\tableno by 1
\relax\ifproofmode\ifforwardreference
\immediate\write1{\noexpand\expandafter\noexpand\def
\noexpand\csname TABLABEL#1\endcsname{\the\chapno.\the\tableno?}}\fi\fi
\global\expandafter\edef\csname TABLABEL#1\endcsname
{\the\chapno.\the\tableno?}\iftablechapternumbers\chapfolio.\fi
\ifproofmode\llap{\hbox{\marginstyle#1
\kern1.2truein}}\relax\fi\the\tableno}

\def\tabref#1{\hbox{\ifundefined{TABLABEL#1}!!!!\ifproofmode%
\ifforwardreference%
\else\write16{ ***Undefined Table Reference #1*** }\fi
\else\write16{ ***Undefined Table Reference #1*** }\fi
\else\edef\LABtt{\getlabel{TABLABEL#1}}%
\def\LABTT{\expandafter\stripchap\LABtt}\iftablechapternumbers%
\chapshow{\LABTT}.\expandafter\stripeq\LABtt%
\else\ifnum\number\LABTT=\chapno\relax\expandafter\stripeq\LABtt%
\else\chapshow{\LABTT}.\expandafter\stripeq\LABtt\fi\fi\fi}%
\ifproofmode\write2{Table#1}\fi}

\def\fig{Figure~}

\newdimen\sectionskip     \sectionskip=20truept	\sectionskip=10truept
\newcount\sectno
\def\section#1#2{\sectno=0 \null\vskip\sectionskip
    \centerline{\chaplabel{#1}.~~{\bf#2}}\nobreak\vskip.03truein
    \noindent\ignorespaces}

\def\advancesectno{\global\advance\sectno by 1}
\def\sectfolio{\number\sectno}
\def\subsection#1{\goodbreak\advancesectno\null\vskip10pt
                  \noindent\chapfolio.~\sectfolio.~{\bf #1}
                  \nobreak\vskip.05truein\noindent\ignorespaces}

\def\uttg#1{\null\vskip.1truein
    \ifproofmode \line{\hfill{\bf Draft}:
    UTTG--{#1}--\number\year}\line{\hfill\today}
    \else \line{\hfill UTTG--{#1}--\number\year}
    \line{\hfill\ifcase\month\or January\or February\or March\or April\or
May\or June
    \or July\or August\or September\or October\or November\or December\fi
    \space\number\year}\fi}

\def\getlabel#1{\csname#1\endcsname}
\def\ifundefined#1{\expandafter\ifx\csname#1\endcsname\relax}
\def\stripchap#1.#2?{#1}
\def\stripeq#1.#2?{#2}

%
\catcode`@=11 
\def\space@ver#1{\let\@sf=\empty\ifmmode#1\else\ifhmode%
\edef\@sf{\spacefactor=\the\spacefactor}\unskip${}#1$\relax\fi\fi}
\newcount\referencecount     \referencecount=0
\newif\ifreferenceopen       \newwrite\referencewrite
\newtoks\rw@toks
\def\refmark#1{\relax[#1]}
\def\refend{\refmark{\number\referencecount}}
\newcount\lastrefsbegincount \lastrefsbegincount=0
\def\refsend{\refmark{\count255=\referencecount%
\advance\count255 by -\lastrefsbegincount%
\ifcase\count255 \number\referencecount%
\or\number\lastrefsbegincount,\number\referencecount%
\else\number\lastrefsbegincount-\number\referencecount\fi}}
\def\refch@ck{\chardef\rw@write=\referencewrite
\ifreferenceopen\else\referenceopentrue
\immediate\openout\referencewrite=referenc.texauxil \fi}
%
{\catcode`\^^M=\active 
  \gdef\obeyendofline{\catcode`\^^M\active \let^^M\ }}%
%
{\catcode`\^^M=\active 
  \gdef\ignoreendofline{\catcode`\^^M=5}}
{\obeyendofline\gdef\rw@start#1{\def\t@st{#1}\ifx\t@st\blankend%
\endgroup\@sf\relax\else\ifx\t@st\bl@nkend\endgroup\@sf\relax%
\else\rw@begin#1
\backtotext
\fi\fi}}
{\obeyendofline\gdef\rw@begin#1
{\def\n@xt{#1}\rw@toks={#1}\relax%
\rw@next}}
\def\blankend{}
{\obeylines\gdef\bl@nkend{
}}
\newif\iffirstrefline  \firstreflinetrue
\def\rwr@teswitch{\ifx\n@xt\blankend\let\n@xt=\rw@begin%
\else\iffirstrefline\global\firstreflinefalse%
\immediate\write\rw@write{\noexpand\obeyendofline\the\rw@toks}%
\let\n@xt=\rw@begin%
\else\ifx\n@xt\rw@@d \def\n@xt{\immediate\write\rw@write{%
\noexpand\ignoreendofline}\endgroup\@sf}%
\else\immediate\write\rw@write{\the\rw@toks}%
\let\n@xt=\rw@begin\fi\fi\fi}
\def\rw@next{\rwr@teswitch\n@xt}
\def\rw@@d{\backtotext} \let\rw@end=\relax
\let\backtotext=\relax

\newdimen\refindent     \refindent=30pt
\def\Textindent#1{\noindent\llap{#1\enspace}\ignorespaces}
\def\refitem#1{\par\hangafter=0 \hangindent=\refindent\Textindent{#1}}
\def\REFNUM#1{\space@ver{}\refch@ck\firstreflinetrue%
\global\advance\referencecount by 1 \xdef#1{\the\referencecount}}
\def\refnum#1{\space@ver{}\refch@ck\firstreflinetrue%
\global\advance\referencecount by 1\xdef#1{\the\referencecount}\refend}

\def\REF#1{\REFNUM#1%
\immediate\write\referencewrite{%
\noexpand\refitem{#1.}}%
\begingroup\obeyendofline\rw@start}
\def\ref{\refnum\?%
\immediate\write\referencewrite{\noexpand\refitem{\?.}}%
\begingroup\obeyendofline\rw@start}
\def\Ref#1{\refnum#1%
\immediate\write\referencewrite{\noexpand\refitem{#1.}}%
\begingroup\obeyendofline\rw@start}
\def\REFS#1{\REFNUM#1\global\lastrefsbegincount=\referencecount%
\immediate\write\referencewrite{\noexpand\refitem{#1.}}%
\begingroup\obeyendofline\rw@start}

\def\cite#1{\refmark#1}
\catcode`@=12 
%
\proofmodefalse
\chapternumberstrue
\forwardreferencetrue
\ifproofmode
\immediate\openout2=allcrossreferfile 
\ifforwardreference\input labelfile
\immediate\openout1=labelfile \fi\fi

\def\subsection#1{\goodbreak\advancesectno\null\vskip10pt
                  \noindent{\it \chapfolio.\sectfolio.~#1}
                  \nobreak\vskip.05truein\noindent\ignorespaces}


%
\input epsf
\nopagenumbers\pageno=-1
\null\vskip-20pt
\rightline{\eightrm hep-th/9703003}\vskip-1pt
\rightline{\eightrm JHU-TIPAC-96028}\vskip-3pt
\rightline{\eightrm TUW-97-05}\vskip-3pt
\rightline{\eightrm UTTG-07-97}\vskip-1pt
\rightline{\eightrm February 1997}
\vskip1truein
\centerline{\seventeenrm The Web of Calabi--Yau Hypersurfaces}
\vskip10pt
\centerline{\seventeenrm in Toric Varieties}
\vskip5pt
\centerline{\seventeenrm }
\vskip.6truein
\centerline{ A.C.~Avram \footnote{*}{Email: alex@physics.utexas.edu},~
             M.~Kreuzer \footnote{**}{Email: kreuzer@tph16.tuwien.ac.at},~ 
             M.~Mandelberg \footnote{\dag}{Email: isaac@bohr.pha.jhu.edu}
             and~
             H.~Skarke \footnote{\ddag}{Email: skarke@zerbina.ph.utexas.edu}}

\vskip20pt
\centerline{(*, \ddag) \it Theory Group, Department of Physics, University of
            Texas at Austin}
\centerline{\it Austin, TX 78712, USA}
\centerline{(**) \it Institut f\"ur Theoretische Physik, 
         Technische Universit\"at Wien}
\centerline{\it         Wiedner Hauptstra\ss e 8--10, A-1040 Wien, AUSTRIA  }
\centerline{ (\dag) \it The Johns Hopkins University, Baltimore, MD, 21218, 
	USA}
\centerline{\it }
\centerline{\it }
\vskip.6truein
\vbox{\centerline{\bf ABSTRACT}
\vskip.2truein
\vbox{\baselineskip=13pt \noindent 
Recent results on duality between string theories and connectedness of their
moduli spaces seem to go a long way toward establishing the uniqueness of an 
underlying theory.
For the large class of \cy\ 3-folds that can be embedded as hypersurfaces in 
toric varieties the proof of mathematical connectedness via singular limits 
is greatly simplified  by using polytopes that are maximal with respect to 
certain single or multiple weight systems.
We identify the multiple weight systems occurring in this approach.
We show that all of the corresponding \cys\ are connected among themselves and 
to the web of CICY's.
This almost completes the proof of connectedness for toric \cy\ hypersurfaces.
}} 
\vfill\eject
\pageno=1\footline={\rm\hfil\folio\hfil}
\baselineskip=17pt plus 1pt minus 1pt

%
\REF{\rReid}{M.~Reid, Math.\ Ann.\ {\bf 278} (1987) 329.}
\REF{\rCDLS}{P.~Candelas, A.M.~Dale, C.A.~L\"utken, R.~Schimmrigk,\hfill\break
       \npb{298}~(1988)~493.}
\REF{\rRolling}{P.~Candelas, P.S.~Green and T.~H\"ubsch,
      \npb{330} (1990) 49.}
\REF{\rAGM}{P.~S.Aspinwall, B.R.~Greene and D.R.~Morrison,\hfil\break
Int.~Math.~Res.~Notices (1993)~ 319, alg-geom/9309007.}
\REF{\rBeast}{T.~H\"ubsch, {\it \cy\ Manifolds--A Bestiary for
       Physicists},\hfil\break (World Scientific, Singapore, 1992).}
\REF{\rTris}{T.~H\"ubsch, Commun.~Math.~Phys.~{\bf 108}~(1987)~291.}
\REF{\rPGTH}{P.~Green, T.~H\"ubsch, Commun.~Math.~Phys.~{\bf 109}~(1987)~99.}
\REF{\rHeC}{A.~He and P.~Candelas, ~Commun.~Math. Phys.~{\bf 135}~193~(1990).}
\REF{\rLS}{M.~Lynker and R.~Schimmrigk, ``Conifold Transitions and
Mirror Symmetries'',\hfill\break hep-th/9511058.}
\REF{\rCLS}{P.~Candelas, M.~Lynker and R.~Schimmrigk, \npb{341}~(1990)~383.}
\REF{\rKS}{A.~Klemm, R.~Schimmrigk, \npb{411}~(1994)~559.}
\REF{\rMaxSkI}{M.~Kreuzer, H.~Skarke, \npb{388}~(1992)~113.}
\REF{\rACJM}{A.C.~Avram, P.~Candelas, D.~Jancic, M.~Mandelberg, 
\npb{465}~(1996)~458.}
\REF{\rCGGK}{T.~Chiang, B.~Greene, M.~Gross and Y.~Kanter, ``Black
Hole Condensation and the Web of Calabi--Yau Manifolds'', hep-th/9511204.}
\REF{\rKSk}{M.Kreuzer, H.Skarke, ``On the Classification of
Reflexive Polyhedra'',\hfil\break 
hep-th/9512204.}
\REF{\rSk}{H.~Skarke, Mod. Phys. Lett. {\bf A11} (1996) 1637,
alg-geom/9603007.}
\REF{\rB}{V.~Batyrev, Duke Math. Journal, Vol 69, No 2, (1993)
      349, ~alg-geom/9310003.}
\REF{\rBKK}{P.~Berglund, S.~Katz and A.~Klemm, ``Mirror Symmetry and
the Moduli Space for Generic Hypersurfaces in Toric Varieties'',
hep-th/9506091.}
\REF{\rStro}{A.~Strominger, \npb{451}~(1995)~97, ~hep-th/9504090.}
\REF{\rGMS}{B.~Greene, D.~R.~Morrison and A.~Strominger, \npb{451}~(1995)~109,
\hfill\break hep-th/9504145.}
\REF{\rCox}{D.~Cox, J. Alg. Geom. {\bf 4} (1995) 17.}
\REF{\rSchoell}{J. Sch\"oll, unpublished.} 
\REF{\rFIB}{A.C. Avran, M. Kreuzer, M. Mandelburg and H. Skarke,
	``Searching for K3 fibrations", hep-th/9610154.}
\REF{\rAsp}{P.~Aspinwall, ``An N=2 Dual Pair and a Phase Transition",
 hep-th/9510142.}
\REF{\rMorii}{D.~R.~Morrison, ``Through the Looking Glass",
 Lecture at CIRM conference, Trento 1994.}
\REF{\rAM}{P.~Aspinwall, D.~R.~Morrison, 
\plb{355} (1995),141.} 
\REF{\rMori}{D.~R.~Morrison,``Mirror Symmetry and Type II Strings",
Proceedings of Trieste Workshop on S Duality and Mirror Symmetry.}
\REF{\rWil}{P.~M.~Wilson, Invent.Math. 107(1992), 561.}
\REF{\rMV}{D.~R.~Morrison, C.~Vafa, ``Compactifications of F-Theory on \cy\
Threefolds II", hep-th/9603161.}
\REF{\rCDK}{P.~Candelas, Xenia~de la Ossa and
S.~Katz, \npb{450}~(1995)~267,\hfil\break hep-th/9412117.}
\REF{\rHaya}{Y.~Hayakawa, ``Degeneration of Calabi--Yau Manifold
With Weil--Peterson  Metric'', ~alg-geom/9507016.}
\REF{\rAs}{P.~Aspinwall, Phys. Lett.{\bf B 371}(1996)231.}
\REF{\rBSV}{M.~Bershadsky, V.~Sadov, C.~Vafa, ``D-Strings on D-Manifolds",
hep-th/9510225.}
\REF{\rKM}{A.~Klemm, P.~Mayr, ``Strong coupling singularities and nonabelian 
gauge symmetry in N=2 string theory", hep-th/9601014.}
\REF{\rKMP}{S.~Katz, D.~R.~Morrison, M.~R.~Plesser, ``Enhanced Gauge Symmetry
in Type II String Theory", hep-th/9601108.}
\REF{\rBKKM}{P.~Berglund, S.~Katz, A.~Klemm, P.~Mayr, ``New Higgs Transitions
between dual N=2 String Models", hep-th/9605154.}
\REF{\rAD}{P.~C.~Argyres, M.~Douglas, \npb{448}~(1995)~93.}
\REF{\rBLS}{I.~Brunner, M.~Lynker, R.~Schimmrigk, ``Unification of M and F
Theory \cy\ Fourfold Vacua", hep-th/9610195.}
\REF{\rAAS}{M. Kreuzer, H. Skarke, ``All abelian symmetries of 
	Landau--Ginzburg potentials", \npb{405}~(1993) 305.}

\section{intro}{Introduction}
The work on the connectedness of the moduli space of \cys\ started with 
\cite{{\rReid,\rCDLS,\rRolling,\rAGM}} where it was noted that different 
components of the moduli space meet along boundaries that correspond to 
singular
manifolds. 
Since M.Reid's conjecture \cite{\rReid} that the parameter space of
3-folds with vanishing first Chern class is connected, different steps have 
been
taken in trying to make it a theorem. 
In \cite{{\rRolling,\rBeast}} the connectedness
of complete intersection \cys\ (CICY's) \cite{{{\rTris}, {\rPGTH}, {\rHeC}, 
{\rLS}}}
was proven. 
The second class of \cys\ that has been constructed is made up of transverse 
hypersurfaces in weighted projective spaces 
\cite{{\rCLS, \rKS, \rMaxSkI}}.
The authors of \cite{{\rKS, \rMaxSkI}} constructed a list of 7555 weight 
vectors 
corresponding to these  varieties. 
This class of hypersurfaces has been shown
to be connected in \cite{{{\rACJM}, {\rCGGK}}} by using toric geometry 
techniques 
which are similar to the ones used in this paper.

Since it was realized that to each dual 
pair of reflexive polyhedra of dimension
four one can associate a family of \cys\ embedded in the corresponding  toric 
variety,
an outstanding problem has been the explicit construction of these polytopes. 
Fortunately the proof of the connectedness of the moduli space of \cy\ 
hypersurfaces
does not require an explicit enumeration of all such families. 
The way toward a solution was paved by refs.
\cite{{\rKSk, \rSk}} which make it possible to construct 
\new a set of maximal \endnew
reflexive polytopes in dimension four that 
contain all others. 
Proving that the respective families of \cys\ are connected would imply 
the connectedness of the whole class of hypersurfaces.

In the classification program presented in \cite{{\rKSk, \rSk}} the 
central role is played by the weight systems with the `{\it interior 
point}' and the `{\it span}' properties 
\new (see section 3)\endnew. 
To each such weight system there corresponds a reflexive polyhedron,
namely the associated maximal Newton polyhedron (MNP). 
The crucial observation of \cite{{\rKSk, \rSk}} is that any 
reflexive polyhedron is a subpolyhedron of one of these or of
certain MNPs defined by more than one weight system or of polyhedra 
that arise upon restriction of an MNP to some sublattice.
In the present work we take another step towards the classification
of reflexive polyhedra by identifying all possible combinations of weight 
systems that are relevant in this context.
We will show that this set is connected;
as we expect the set of polyhedra coming from sublattices to be rather small,
this means that we have done most of the work necessary to show the 
connectedness of all toric Calabi--Yau hypersurfaces.

The correspondence between \cys\ and reflexive polyhedra has been described in 
the work of Batyrev  \cite{{\rB}}. 
We consider varieties given by the zero locus of a polynomial $p$
that contains all monomials  satisfying certain constraints. 
The monomials are in one-to-one correspondence with the integer points of 
the polyhedron $\D$. 
If $\Delta$ has a property termed reflexivity, then 
there is a family ${\cal M}_{(\Delta, \D^*)}$ of \cy\ hypersurfaces  
$p = 0$ in the
toric variety ${\cal V}_{\S_\Delta^*}$ defined by a fan over (some 
triangulation of) the dual polytope $\D^*$ \cite{\rB}.

The interplay between the analytic properties of \cys\ and the geometry of 
reflexive polyhedra will be used to prove that the moduli space of these 
varieties is connected.
Generalizing the concept of ${\cal M}_{(\Delta, \D^*)}$,
we associate a family of hypersurfaces ${\cal M}_{(\D_1, \D_2^*)}$ even to 
non--dual pairs of reflexive polyhedra $(\D_1, \D_2^*)$ .
If  $\D$ contains a reflexive subpolyhedron $\delta$ then 
${\cal M}_{(\d, \d^*)}$ is birational to the subfamily 
${\cal M}_{(\d, \D^*)}$ of ${\cal M}_{(\Delta, \D^*)}$. 
The moduli spaces of ${\cal M}_{(\Delta, \D^*)}$ and  ${\cal M}_{(\d, \d^*)}$
overlap on the subfamily ${\cal M}_{(\d, \D^*)}$ \cite{\rBKK}.

We emphasize 
that the singular manifolds where different regions of 
the moduli space touch have, in many cases, singularities different
from the conifold type analyzed in \cite{{{\rStro}, {\rGMS}}} and 
the physics associated with these spaces is not  completely understood.
The problem of determining the low energy effective theory of the 
Type II string compactified on an arbitrary singular variety, and describing 
the associated extremal transition, remains open.
Current methods would demand that such an analysis be done on a case by case 
basis. 
We do not attempt this here.

In \SS2 we briefly review the basics of \cy\ embeddings in toric varieties.
After a summary of the methods and results of refs. \cite{{\rKSk, \rSk}},
we present our results on the classification of the combinations of
weight systems in \SS3.
The relevance of polyhedra being contained in one another is presented in \SS4
after which we illustrate the  method with a two dimensional example in \SS5.
The main steps of the computation that proves the connectedness
are overviewed in \SS6.
We present concluding remarks in \SS7.
Two tables give information on data of some new 
\cy\ hypersurfaces related to combinations of weight systems.

\section{CYtoric}{Calabi--Yau hypersurfaces in toric varieties}
Probably the largest class of \cys\ explicitly constructed up to  now 
is represented by hypersurfaces in toric varieties.
We first review some of the basic principles underlying 
the interplay between reflexive polyhedra, toric varieties and the sections 
of the anticanonical bundle (sheaf) over these spaces.

We start with a reflexive polyhedron $\D$ defined by its vertices belonging 
to  a\ lattice $M$.
We recall that a reflexive polyhedron satisfies the following conditions: 
(i) it has integer vertices; 
(ii) it has only one interior point; 
(iii) the equation of any face of codimension $1$ can be written in the form
$\, c_1 x^1 + \dots +c_n x^n =1$, where the $c_i$'s are integers with no
common divisor, the $x^i$'s are coordinates on $M$, and $n = \hbox{ rank }(M)$.
We define the polytope $\D^*$ to be dual to $\D$, i.e.
$$ \Delta^* \, = \, \{ {\bf y} \in N_\IR \mid 
\langle {\bf y},{\bf x} \rangle \ge -1 \, , \hbox{ for all }
{\bf x} \in \D \},
\eqlabel{npoly}$$
where $N_\IR=N \otimes_{\IZ} \IR$ is the real extension of 
$N \, = \, \hbox{Hom}({M}, \IZ)$. 
Note that $\D^*$ is itself reflexive.

The integer points of $\D^* \cap N$ define the $1$-dimensional cones 
$\{ {\bf v}_1 ,\dots ,{\bf v}_N \}=\S^1_{\D^*}$
of the fan $\Sigma_{\D^*}$, which we assume to correspond to a maximal
triangulation of the fan over the faces of $\D^*$.
The $1$-dimensional cones span the vector space $N_\IR$ and satisfy 
relations of linear dependence
$$ \sum_l \, k_j^l {\bf v}_l = 0 \,, \quad k_j^l \ge 0.
\eqlabel{lindep}$$
%
Following Cox \cite{\rCox},
we can build a variety ${\cal V}_{\Sigma_{\D^*}}$ as the space 
$\IC^N \setminus Z_{\Sigma_{\D^*}}$ modulo the action of a group which is the
product of a finite group and 
the torus $(\IC^*)^{N-n}$. The action of the torus is defined by:
$$ (z_1, \dots ,z_N) \sim (\lambda^{k^1_j} z_1, \dots, \lambda^{k^N_j} z_N) \,,
\quad j=1,\dots,N-n.
\eqlabel{scaling}$$
$Z_{\Sigma_{\D^*}}$ is an exceptional subset of $\IC^N$ defined as
$$ Z_{\Sigma_{\D^*}} \, = \, \bigcup_{\cal I} \{ (z_1,\dots,z_N) \mid z_i =0
\hbox{ for all }  i \in {\cal I} \},
\eqlabel{exset}$$
where the union is taken over all index sets ${\cal I} = \{ i_1,\dots,i_k \} $
such that $\{{\bf v}_{i_1}, \dots,{\bf v}_{i_k} \}$ do not belong to the 
same maximal cone in $\Sigma_{\D^*}$.
This depends explicitly on which triangulation of the fan over $\D^*$
we have chosen. 
The elements of $\S^1_{\D^*}$ are in one-to-one correspondence with T-invariant
divisors $D_{{\bf v}_i}$ on ${\cal V}_{\Sigma_{\D^*}}$.
Knowing the embedding toric variety, ${\cal V}_{\Sigma_{\D^*}}$,
we want to find the space of sections of the anticanonical sheaf.
According to Batyrev \cite{\rB}, this is given in terms of the polytope $\D$:
The points of $\Delta \cap M$ are in one-to-one correspondence 
with monomials in the homogeneous coordinates $z_i$.
\del
(sections of the anticanonical sheaf over ${\cal V}_{\Sigma_{\D^*}}$).
$\Delta$ also corresponds to the space of global sections of the T-Cartier 
divisor $D=\sum D_{{\bf v}_i}$.
\enddel
A general polynomial determining a section of the anticanonical sheaf
in ${\cal V}_{\Sigma_{\D^*}}$ is given by
$$ p = \sum_{{\bf x} \in \Delta \cap M} c_{\bf x} \prod_{l=1}^N 
   z_l^{\langle {\bf v}_l, {\bf x} \rangle + 1}.
\eqlabel{pol}$$
The $c_{\bf x}$ parametrize a family ${\cal M}_\D$ of Calabi-Yau
hypersurfaces defined as the zero loci of $p$.

 \def\ip{\hbox{\bf 1}} 
\def\ipo{\hbox{\bf 0}}	
\def\<{\langle } \def\>{\rangle }
\def\cont#1{\mathop{\vtop{\ialign{##\crcr $\hfil\displaystyle
                     {#1}\hfil$\crcr\noalign{\kern3 pt \nointerlineskip}
                     $\bracelu\leaders\vrule\hfill\leaders\vrule\hfill
                     \braceru$\crcr\noalign{\kern3 pt }}}}\limits}

\section{MP}{Minimal polyhedra and combinations of weight systems}

Having in view that reflexive polyhedra encode properties of families
of \cy\ hypersurfaces in toric varieties, we now show how one might
look for all inequivalent reflexive polyhedra. 
To this end we review and extend concepts and results from refs.
\cite{{\rKSk, \rSk}}.
\del
Consider a dual pair of lattices $N$ and $M$, each isomorphic
to $\IZ^n$, and their real extensions $N_\IR$ and $M_\IR$.
\enddel
We recall that any polytope in $M_\IR$ with the origin in its interior can be 
described by a set of inequalities $\<{\bf n}_i,{\bf x}\>\ge -1$ with 
${\bf n}_i\in N_\IR$.
A complete and non-redundant description is provided if the set
of ${\bf n}_i$ corresponds to the set of vertices $V_i$ of the dual polytope.
In particular, reflexivity of the polytope $\D$ implies that
$V_i\in N$.
Of course $\D$ is a subset of any polyhedron defined only by
a subset of the above mentioned inequalities.

The main idea of the classification program for reflexive
polyhedra initiated in \cite{\rKSk} was the introduction of
so-called minimal polyhedra. These are polytopes defined by
a collection of inequalities such that a polyhedron defined
by any strict subset of this collection would be unbounded.
In terms of the dual space this has the following meaning:
if $Q$ is a minimal polyhedron (with respect to hyperplanes), 
then $Q^*$ is a polytope in $N_\IR$
whose vertices are the $V_i$ involved in the description
of $Q$, in such a way that the convex hull of any strict subset 
of these $V_i$ does not have the origin $\ipo_N$ of $N_\IR$ in its 
interior (i.e., $Q^*$ is minimal with respect to vertices).
The possible shapes of these objects were classified in \cite{\rKSk}.
By considering triangulations, 
it is more or less easy to see that any minimal polyhedron
must be the convex hull of the set of vertices of some simplices
(possibly of lower dimension) which all contain $\ipo_N$
in their interiors.
The fact that subsets of such a set of simplices give rise
to lower dimensional minimal polyhedra makes an iterative
construction of all minimal polyhedra for rising
dimensions possible. 
This construction was explained  and applied to the cases
with $n\le 4$ in ref. \cite{\rKSk},
with the following results on the structure of $Q^*$.

In one dimension the only possibility is a line segment
(`1-simplex') $V_1V_2$ with $\ipo_N$ in its interior.
In two dimensions there are the triangle $V_1V_2V_3$ and
the parallelogram that is 
the convex hull of $V_1,V_2,V_1',V_2'$ such that $V_1V_2$
and $V_1'V_2'$ are 1-simplices with $\ipo_N$ in their (1-d)
interiors. 
The fact that simplices with $\ipo_N$ in their interiors
are the building blocks of minimal polytopes remains
valid in arbitrary dimensions.
Representing simplices by the numbers of their vertices,
the results on the classification of minimal polyhedra can
be summarized in the following way:

\noindent
n=1: ~~~~ 2.
\vskip3pt\noindent
n=2: ~~~~ 3; ~~~ 2+2.
\vskip3pt\noindent
n=3: ~~~~ 4; ~~~ 3+2, ~ $\cont{\hbox{3+3}}$; ~~~ 2+2+2.
\vskip3pt\noindent
n=4: ~~~~ 5; 
	~~~ 4+2, ~ 3+3, ~ $\cont{\hbox{4+3}}$, 	~ $\cont{\cont{\hbox{4+4}}}$;
	~~~ 3+2+2, ~ $\cont{\hbox{3+3}}$+2, ~ 
	$\cont{\hbox{3+3}\hskip -3pt}\hskip -2pt\cont{\hbox{~+3}}$
                ~~~ 2+2+2+2.
\hfil\break
The underlining symbols indicate common vertices of simplices.
For example, $\cont{\hbox{3+3}}$ means the convex hull of 
$V_1,V_2,V_3,V_2',V_3'$,
where $V_1V_2V_3$ and $V_1V_2'V_3'$ are triangles with $\ipo_N$ in their 
interiors, sharing the vertex $V_1$. 
In a similar way $\cont{\cont{\hbox{4+4}}}$ stands for two tetrahedra 
sharing two vertices and $\cont{\hbox{3+3}\!\!}\!\cont{\hbox{~+3}}$
stands for three triangles sharing a single vertex.
Denoting by $k$ the number of vertices of $Q^*$, the number
of simplices in this description is always $k-n$.

Each of the simplices occurring above may be used to define
a weight system $(q_i)$ that corresponds to the barycentric
coordinates of $\ipo_N$ with respect to the vertices of the
simplex $\sum q_i=1,\;\sum q_i V_i=\ipo_N$.
These weight systems are the major tool for a convenient description
of our original minimal polyhedra $Q\subset M_\IR$.
Recall that the most symmetric description of an $n$ dimensional
standard simplex is as the convex hull of base vectors in $\IR^{n+1}$
or, equivalently, the intersection of the positive hyperoctant in
$\IR^{n+1}$ with the hyperplane $\sum_{i=1}^{n+1}x^i=1$.
In a similar way, if $Q^*$ is a simplex giving rise to
a weight system $(q_i)$, then $Q$ may be represented as the intersection 
of the positive octant in $\G^{n+1}_\IR\simeq\IR^{n+1}$ with the 
hyperplane $\sum_{i=1}^{n+1}q_ix^i=1$.
For rational weights the latter equation defines an $n$ dimensional
sublattice $\G^n$ of $\G^{n+1}\simeq\IZ^{n+1}$. 
The $M$ lattice can be identified with a sublattice of $\G^n$, 
and $\ipo_M$ corresponds to $(1,\cdots,1)=:\ip$.

A similar prescription also works for minimal polyhedra that
are not simplices. If $Q^*$ has $k>n+1$ vertices, we may
represent $Q$ as the intersection of the positive octant in 
$\G^k_\IR\simeq\IR^k$ with $k-n$ hyperplanes of the type
$\sum_{i=1}^kq_ix^i=1$.
For example, if $Q^*$ is of the type $3+3$, the 
4-d polytope $Q$ is given by the positive octant in $\IR^6$
intersected with two hyperplanes determined by the weight
systems $(q_1,q_2,q_3,0,0,0)$ and $(0,0,0,q_1',q_2',q_3')$;
if $Q^*$ is of the type $\cont{\hbox{3+3}}$, the 
3-d polytope $Q$ is given by the positive octant in $\IR^5$
intersected with two hyperplanes determined by the weight
systems $(q_1,q_2,q_3,0,0)$ and $(q_1',0,0,q_2',q_3')$.
Note that if the simplices in $Q^*$ have no vertex in common,
then $Q$ is just the product of the simplices defined by
the weight systems.

Given a weight system or a combination of weight systems,
we call the convex hull of $Q\cap M$ the maximal Newton 
polyhedron $\D_{\rm max}$ corresponding to the (combination of) 
weight system(s).
The weight systems that play a role for the construction of
reflexive polyhedra are just those with the property that their 
maximal Newton polyhedra (with respect to $M=\G^n$) have $\ip$ in 
their interiors.
It is easy to see that a combination of weight systems
can have this property only if each of the weight systems has it.
If the simplices in $Q^*$ have no common points, the converse
is also easily seen to be true.
The weight systems with up to 5 weights which have this property
have been classified in \cite{\rSk}.
In the same paper it was also shown that for dimension $n\le 4$
any maximal Newton polyhedron (whether with respect to a single
weight system or a combination) with the interior point property
is actually reflexive.
Yet we do not need all of these weight systems for the classification
program. 
Remembering that we assumed the $V_i$ to be vertices of $\D^*$,
we see that the hyperplanes in $M_\IR$ dual to the $V_i$
should carry facets of $\D$. 
This can only happen if these hyperplanes are affinely spanned by
points of $\D_{\rm max}$.
Again, for a combination of weight systems to have this property,
it is a necessary condition that each of the systems involved
has it; for direct products it is also sufficient.

Let us briefly summarize the existing classification methods and 
results on weight systems
whose maximal Newton polyhedra have $\ip$ in their interiors.
The basic idea of the scheme of \cite{\rSk} is to reconstruct
a weight system from the integer points it allows.
Given a set $S=\{{\bf x}_{(j)}\}$ of points in $\IZ^{n+1}_{\ge 0}$ 
containing $\ip$,
it is relatively easy to find out whether this set allows
any weight system $(\tilde q_i)$ with  $\sum_{i=1}^{n+1}\tilde q_ix^i_{(j)}=1$ 
for all $j$ and, if it exists, to find such a system.
If $(\tilde q_i)$ has the interior point property, we can
add it to our list of allowed systems.
If we assume that there is another system $(q_i)$ compatible with 
all ${\bf x}_{(j)}\in S$, then we may see the equation 
$\sum_{i=1}^{n+1}\tilde q_ix^i=1$ 
as the equation of a hyperplane through $\ip$ in the space
$\G^n$ defined by $(q_i)$.
For $\D_{\rm max}$ to have $\ip$ in the interior, there
must be a point `below' this hyperplane, i.e. a point ${\bf x}_{\rm new}$
fulfilling $\sum_{i=1}^{n+1}\tilde q_ix^i_{\rm new}<1$.
The number of such points is finite, so we may apply the same 
considerations to all of the sets $S'=S\cup\{{\bf x}_{\rm new}\}$.
Taking $S=\{\ip\}$ as our starting point, it is possible
to generate in this way all allowed weight systems because 
the new points are always affinely independent of the others
(in $\IZ^{n+1}$), implying that there is no need to go further than
to the $n^{\rm th}$ recursion level in the algorithm.

Applying this algorithm to the cases with $n+1\le 5$ yields the
following results:
There is one interior point property system with 2 weights, 
namely (1/2, 1/2).
There are three such systems with 3 weights: 
(1/3, 1/3, 1/3), (1/2, 1/4, 1/4) and (1/2, 1/3, 1/6). 
All of these systems have the span property. 
With four weights there are 95 such systems, 58 of them having the
span property, and with five weights there are 184,026 systems
among which 38,730 have the span property.

There are still three more steps to be taken for a complete
classification of all reflexive polyhedra:

\noindent
(1) The identification of all combinations of weight systems
that have both the interior point and the span property.\hfill\break
(2) The identification of allowed sublattices such that the
above properties are preserved.\hfill\break
(3) The enumeration of all reflexive subpolyhedra of the maximal 
polyhedra.

Step (1) has been taken now. 
For direct products it is trivial because
they have the interior point and the span property
if and only if every factor has both properties.
For other types of combinations we applied the following scheme:
We first took all combinations of spanning weight systems.
To write a computer program that creates  all inequivalent combinations 
while at the same time avoiding any redundancy is an easy
exercise in elementary combinatorics.
Then we checked the combined system for the span property, using 
software developed for \cite{\rSk}.
In a last step we then checked for the reflexivity
of the maximal Newton polyhedron (which, by the theorem of \cite{\rSk},
is equivalent to the interior point property).
To this end we used software developed for \cite{\rACJM}
and independently checked the results with a different package
\cite{\rSchoell}.

We obtained the following results: For $n=3$ we got, in addition to the
58 single weight systems with 4 weights, the obvious 3 combinations
of the type 3+2, 17 combinations of the type $\cont{\hbox{3+3}}$
and of course the single combination 2+2+2.
Altogether we have obtained 79 polyhedra in three
dimensions which contain all reflexive 3d polyhedra.
For $n=4$ the numbers are:

\noindent
5: ~38730; ~~~ 4+2: ~58, ~~ 3+3: ~6, ~~ $\cont{\hbox{4+3}}$: ~727, 
~ $\cont{\cont{\hbox{4+4}}}$: ~6365;
\vskip-8pt\noindent
3+2+2: ~3, ~~ $\cont{\hbox{3+3}}$+2: ~17, ~~ 
$\cont{\hbox{3+3}\!\!}\!\cont{\hbox{~+3}}$: ~36;  ~~~ 2+2+2+2: ~1.

\noindent
The total number of four dimensional polyhedra obtained in this
way is 45,943. 
\new
In table~1 we list the 426 new spectra that multiple weight systems yield
in addition to the 10238 spectra%
\Footnote{These can be found at 
	http://tph.tuwien.ac.at/{$^\sim$}kreuzer/CY}
that we obtained for single weight systems \cite{\rFIB}. 
In table~2 we detail the results for the various types of 
combinations and give, in the last two colums, the number of new spectra 
(\#new') and new mirror pairs of spectra (\#sym') when also the orbifold 
results in weighted $\IP^4$ are taken into account. 
Altogether we thus obtained 
100 mirror pairs of spectra for which no
reflexive polytopes were previously known.
\endnew

Step (2) requires more work and its results will be reported elsewhere. 
It is likely that most if not all polyhedra defined by sublattices are
subpolyhedra of the ones defined by the original lattices so we expect 
very few cases relevant to the connectedness proof are left out at this
stage. Step (3) is not necessary for the present work which is
why the number of polytopes that had to be effectively connected is manageably
small.

\section{conn}{Connecting different regions of the moduli space}

As mentioned in the introduction the final goal is to show that the moduli 
space of  Calabi--Yau hypersurfaces (CYH) of dimension three in toric
varieties is connected. 
Because of the one-to-one correspondence between subfamilies of 
CYH (with defined Hodge numbers) 
$\cal M$ and pairs of reflexive polyhedra $(\Delta, \D^*)$, the task of 
showing 
that ${\cal M }_1$ is connected to ${\cal M}_2$ is 
reduced to that of showing there are 
$\{ \Delta_a, \Delta_b, \dots , \Delta_z \}$ 
such that
$$\Delta_1 \supset  \subset \Delta_a \supset  \subset \Delta_b \supset  
\subset \cdots \Delta_z \supset \subset \Delta_2 \, ,
\eqlabel{inclusion}$$
where $\Delta_a \supset  \subset \Delta_b $ means that either 
$\Delta_a  \subset \Delta_b$
or $\Delta_a \supset \Delta_b$. 
Let us examine the meaning of \eqref{inclusion}.
Assume that  $\Delta_a  \subset \Delta_b$. 
Then we have $\D^*_a \supset \D^*_b$ and the fan $\Sigma_{\D^*_a}$ has more 
\hbox{1-dimensional} cones (collectively labeled by 
$\Sigma_{\D^*_a}^1$) than $\Sigma_{\D^*_b}$. 
Then ${\cal V}_{\Sigma_{\D^*_b}}$
can be deformed into ${\cal V}_{\Sigma_{\D^*_a}}$ by refining 
$\Sigma_{\D^*_b}$,
that is, by ``blowing up'' ${\cal V}_{\Sigma_{\D^*_b}}$.
\del
The subfamily 
${\cal M}_b^{\sharp}$ described in terms of monomials associated with integer 
points in $\Delta_a \cap \Delta_b$ will be smoothed out.
\enddel
Hence we have a proper birational morphism of toric varieties
$ \ca{V}_{\Sigma_{\D^*_a}}
\longrightarrow  \ca{V}_{\Sigma_{\D^*_b}}$
which is biholomorphic 
everywhere but on the exceptional sets associated 
with the divisors in $\Sigma^1_{\D^*_a} \backslash \Sigma^1_{\D^*_b}$. 
\del
According to \cite{\rCDK} the pullback of ${\cal M}^{\sharp}_b$ to 
$\ca{V}_{(\Sigma_a,\D^*_a)}$ is 
\enddel
The divisors introduced in 
${\cal V}_{\Sigma_{\D^*_b}}$ by this procedure will sometimes 
also intersect 
the family of \cy\ hypersurfaces ${\cal M}_b$. 
Returning to \eqref{inclusion} assume for simplicity that a segment of the 
chain looks like this:
$$\Delta_b \supset \Delta_a \subset \Delta_c
\eqlabel{inclusion2}.$$
This tells us that there are families of possibly 
singular hypersurfaces in both 
${\cal V}_{\Sigma_{\D^*_b}}$ and ${\cal V}_{\Sigma_{\D^*_c}}$,
${\cal M}_b^{\sharp}$ and ${\cal M}_c^{\sharp}$ respectively,
that can be blown up to become isomorphic to ${\cal M}_a$.
It is conceivable that the complex structure deformations of ${\cal M}_b$
$({\cal M}_c)$ will result in a smooth manifold. In such a case we won't deal
with an extremal transition \cite{{\rAsp,\rMorii}}. We do not change 
topologies.
The T-divisors in $\Sigma^1_{\D^*_a} \backslash \Sigma^1_{\D^*_{b(c)}}$
do not intersect the family ${\cal M}_{b(c)}^{\sharp}$ which is thus 
isomorphic with ${\cal M}_a$. 
We may express this as a diagram:
$$
\matrix{\ca{M}_b:&~~~~~~\, (~\D_b,\D^*_b~)\cr
        ~~~~~~~~~&\downarrow\cr
        \ca{M}_b^\sharp:&~~~~~~\,(~\D_a,\D^*_b~)\cr
         ~~~~~~~~~&~~~~~~~~~~\, \downarrow\cr
         \ca{M}_a:&~~~~~~\,(~\D_a,\D^*_a~)\cr
         ~~~~~~~~~&~~~~~~~~~~\, \downarrow\cr
         \ca{M}_c^\sharp:&~~~~~~\,(~\D_a,\D^*_c~)\cr
         ~~~~~~~~~&\downarrow\cr
         \ca{M}_c:&~~~~~~\,(~\D_c,\D^*_c~)\cr}\eqlabel{diagram}$$
We start with the family $\ca{M}_b$ whose members generically represent smooth 
surfaces and specialize the polynomial to obtain the 
subfamily $\ca{M}_b^\sharp$ which can be 
resolved into $\ca{M}_a$ by blowing up
${\cal V}_{\Sigma_{\D^*_b}}$ into ${\cal V}_{\Sigma_{\D^*_a}}$. 
In a next step we blow down some other divisors of 
${\cal V}_{\Sigma_{\D^*_a}}$, thereby contracting it to 
${\cal V}_{\Sigma_{\D^*_c}}$.
This takes the family $\ca{M}_a$ to $\ca{M}_c^\sharp\subset\ca{M}_c$.
{}From there, we can reach any point in the moduli space of $\ca{M}_c$
by smoothly varying the coefficients $c_{\bf x}$ defining a monomial
in $\ca{M}_c$.

\section{ex}{A lower dimensional example}

We want to illustrate the theory with a 2-dimensional example \Footnote{A 
4-d example was discussed in \cite{\rACJM}, mostly from the complex 
structure point of view.}. 
We hope 
that by its simplicity it will make the procedure of the previous section 
more understandable.

Consider the triplet of reflexive polyhedra with their respective 
fans represented by dashed lines in Figure 1.
\vskip .1in
\hskip 1in
\epsfxsize 3.5in
\epsfbox {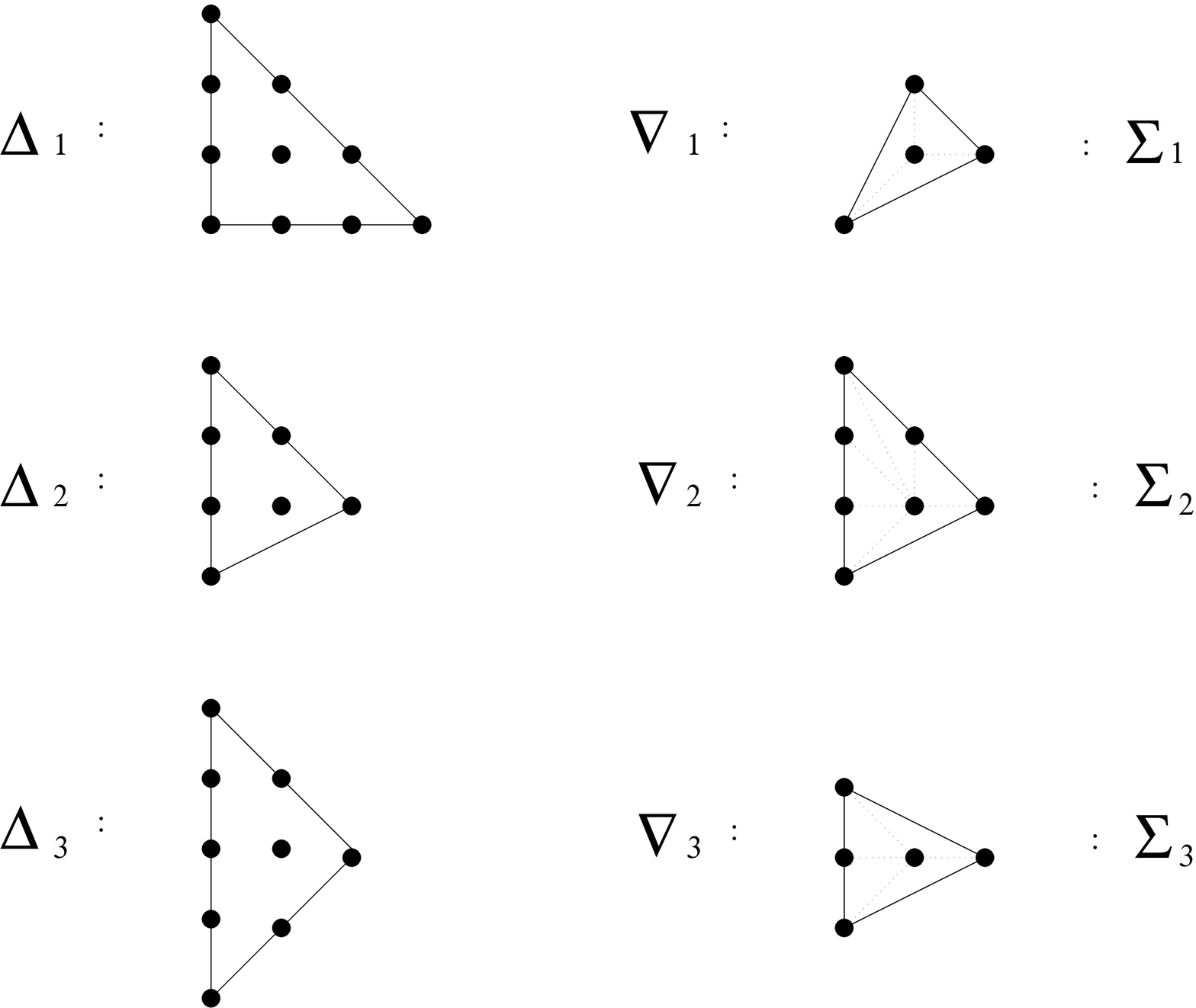}
\vskip .1in
\centerline{\fig1:~{Three pairs of reflexive polyhedra and 
	the associated fans.}}
\vskip .2in
\noindent
By the methods of Section 2 we see that $\Delta_1^*$ corresponds to 
$\IP^2_{[1,1,1]}$, 
while the general polynomial that describes a CYH is 
$$p_1 = x_1^3 + x_2^3 + x_3^3 + x_1x_2x_3 + \cdots$$
$\D^*_2$ has 7 points which determine 6 1-dimensional cones in 
$\Sigma^1_{\D^*_2}$:
$$\{ (1,0), (-1,0), (0,1), (-1,-1), (-1,1), (-1,2) \}$$
which we label ${\bf v}_1,\ldots,{\bf v}_6$.
The general polynomial will be expressed in terms of 6 variables and will 
contain 7 monomials
$$p_2 = z_1^2z_3 + z_2^2z_3z_4^2z_5^2z_6^2 + z_1z_2z_3^2z_5^2z_6^3 + \cdots$$
The $(\IC^*)^4$-action is read from the relations of linear dependence 
$${\bf v}_1 + {\bf v}_2 = 2{\bf v}_1 + {\bf v}_4 + {\bf v}_5 
= {\bf v}_1 + {\bf v}_3 + {\bf v}_4 = 3{\bf v}_1 + 2{\bf v}_4 + {\bf v}_6
= 0$$ 
between elements of $\Sigma^1_{\D^*_2}$ and is given by
$$(z_1, z_2, z_3, z_4, z_5, z_6) \longrightarrow 
(\lambda \mu^2 \nu \rho^3 z_1, 
\lambda z_2, \nu z_3, \mu \nu \rho^2 z_4, \mu z_5, \rho z_6).$$
We can define the following birational map between $\IP^2_{[1,2,3]}$ and 
${\cal V}_{\Sigma_{\D^*_2}}$ 
$$\eqalign{z_1^2z_3 &= y_3^2 \cr
          z_2^2z_4^3z_5 &= y_2^3 \cr
          z_2^2z_3^3z_5^4z_6^6 &= y_1^6}
\eqlabel{birat}$$
The same map is also birational between hypersurfaces in 
${\cal V}_{\Sigma_{\D^*_2}}$ given 
by $p_2 = 0$ and hypersurfaces in $\IP^2_{[1,2,3]}$ given by the zeroes of
$$p_2^{\sharp} = y_1^6 + y_2^3 + y_3^2 + y_1y_2y_3 + \cdots$$
\del
Even though the map is not an isomorphism, by abuse of language we sometimes 
say that the torus can be embedded in $\IP_{[1,2,3]}$. 
\enddel
This is a dimensionally reduced example 
of a \cy\ embedded in a toric variety described by a single weight vector:
$(1,2,3)$.

Before going any further we warn the reader that in our 2-dimensional 
example no 
singularities will be encountered by specializing the defining polynomial.
Singularities may be encountered in dimension 3 and higher.
Of course all 1-dimensional \cys\ are tori and all 2-d \cys\ are 
K3 surfaces, so a transition can involve a change in Hodge 
numbers only if the ambient space has a dimension of 4 or more.  

We can specialize the polynomial $p_1$ of the polyhedron
$\Delta_1$ to $p_1^{\sharp}$ by dropping the monomials $x_1^3$, $x_1^2x_2,$ 
$ x_1x_2^2$.
For $p_1^{\sharp} = 0$ to become isomorphic with
$p_2 =0$ in ${\cal V}_{\Sigma_{\D^*_2}}$ we have to blow up
${\cal V}_{\Sigma_{\D^*_1}}$.
\vskip .2in
\hskip 2in
\epsfxsize .7in
\epsfbox {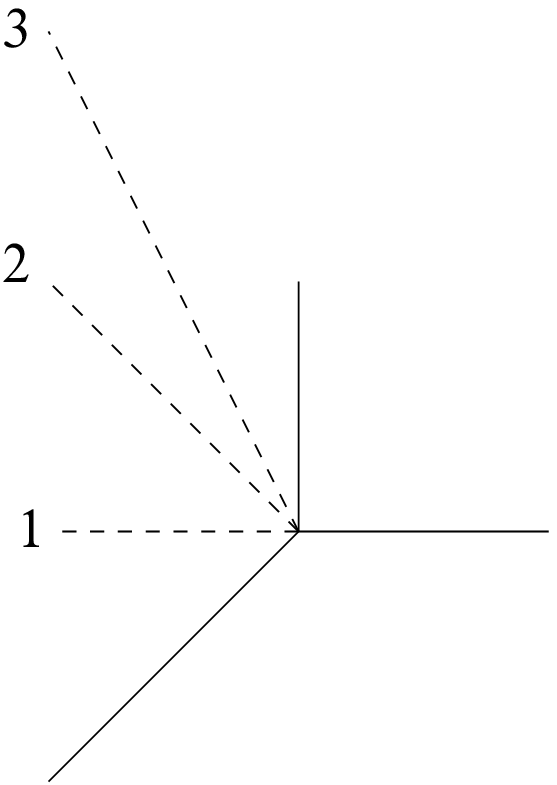}
\vskip .1in
\centerline{\fig2:~{$\S_2$ is a refinement of $\S_1$.}}
\vskip .1in
\noindent
Only one patch of $\IP^2_{[1,1,1]}$ will be affected in the process.
The blow up in this case may  be done ``one divisor at a time''
since there are reflexive polyhedra interpolating between $\D_1$ and
$\D_2$ at each step.
\vskip .1in
\hskip 1in
\epsfxsize 3.5in
\epsfbox {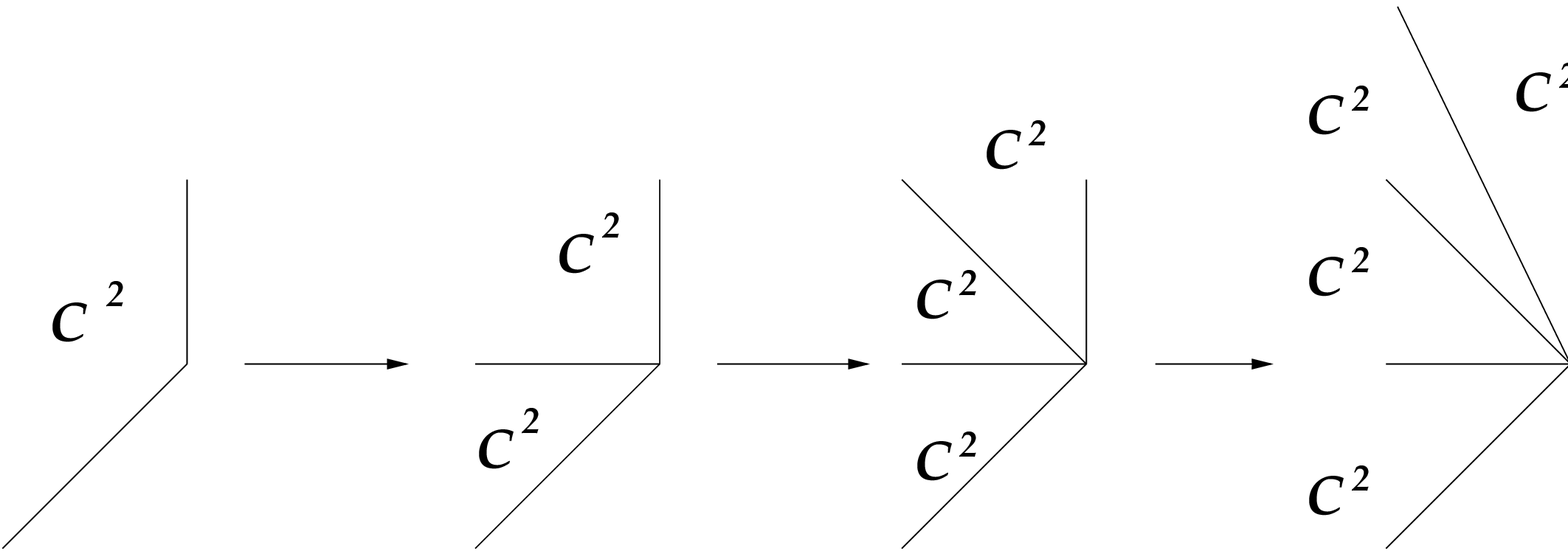}
\vskip .1in
\fig3:~{The successive blow-up of one of the smooth patches of $\IP^2$}\hfill
\vskip .1in
\noindent
Since $\Delta_2$ is also contained in $\Delta_3$, by the same reasoning we 
show that the families
$\ca{M}_{\Delta_2}$ and $\ca{M}_{\Delta_3}$ are connected. 
There is thus a continuous path 
between $\ca{M}_{\Delta_1}$ and $\ca{M}_{\Delta_3}$.
$$\ca{M}_{\Delta_1} \longrightarrow \ca{M}_{\Delta_1}^\sharp \longrightarrow
  \ca{M}_{\Delta_2} \longrightarrow \ca{M}_{\Delta_3}^\sharp \longrightarrow
  \ca{M}_{\Delta_3} $$

\section{comp}{The Computation}

As we have seen in the previous Sections, the moduli spaces of two 
Calabi--Yau varieties defined by reflexive 
polyhedra $\D_1$ and $\D_2$ are 
connected if $\D_1\subset \D_2$ or $\D_2\subset \D_1$.
By abuse of language, we will henceforth speak of
``polyhedra being connected'' instead of ``moduli spaces of Calabi--Yau 
varieties defined by reflexive polyhedra being connected''.
More generally, $\D_1$ and $\D_k$ are called 
connected if there are reflexive polyhedra $\D_2,\cdots,\D_{k-1}$
such that $\D_i$ and $\D_{i+1}$ are connected in the above sense
for $i=1,\cdots,k-1$.
The following statement, which is an immediate consequence of
the results of \cite{{\rKSk,\rSk}} reviewed in Section 3, is the
central point in our scheme for showing the connectedness of
the moduli space.

{\it As any reflexive polyhedron is a subpolyhedron of some
maximal Newton polyhedron defined by a weight system or combination
of weight systems with the interior point and span property,
perhaps with respect to some sublattice, and as these are always
reflexive for $n\le 4$, showing the connectedness
of these maximal Newton polyhedra is sufficient for showing the
connectedness of all reflexive polyhedra.}

In the present work we provide a big step towards this goal by showing
the connectedness of all maximal Newton polyhedra with respect to
the maximal lattice $\G^n$, leaving the classification of sublattices
and the proof of their connectedness to future work.

For $n=2$ we can establish the connectedness of all reflexive polyhedra
(even those obtained from sublattices) by noting that the set of
maximal Newton polyhedra corresponding to a single weight system
is just given by $\D_1$, $\D_2$ and $\D_3$ of Fig. 1.
In addition there is one MNP $\D_4$ coming from the only combination of
weight systems possible for $n=2$.
It can be represented as the square $|x_1|\le 1$, $|x_2|\le 1$
in $M_{\IR}$ with $M\simeq \IZ^2$.
$\D_1$ and $\D_3$ allow for sublattices (w.r.t. $\IZ_3$ and $\IZ_2$,
respectively). 
The reductions of these polyhedra are isomorphic to $\D_1^*$ 
and $\D_3^*$.
The square $\D_4$ also admits a sublattice $\IZ_2$, given by $x_1=x_2$ mod 2, 
and again the reduction to the sublattice is isomorphic to the
dual ($\D_4^*$, in this case).
Perhaps the simplest way of establishing connectedness is by noticing
that any of our maximal polyhedra except $\D_4^*$ contains either 
$\D_1^*$ or $\D_3^*$ 
and that these two polygons are both contained in $\D_2\simeq \D_2^*$
while $\D_4^*$ is contained in most of
the other maximal polyhedra.

For $n=3$, the maximal Newton polyhedra corresponding to the
types 3+2 and 2+2+2 are just prisms of height two over the 
maximal Newton polyhedra for $n=2$, so their connectedness is
established by the connectedness for $n=2$. In order to 
show that they belong to the ``web of $n=3$ polyhedra'', 
we only have to show that one of them is connected with one
of the other polyhedra. Showing the connectedness of the
58 polyhedra defined by a single weight system and the 17
polyhedra of type $\cont{\hbox{3+3}}$ of course requires more work
and has to be done by computer. 
While it is highly probable that K3 hypersurfaces are connected at the level
of polyhedra, we did not attempt to prove this because the connectedness 
of the whole class of K3 hypersurfaces follows trivially from the fact
that this class contains only a single family, anyway.

In the 4-dimensional case which we are most interested in,
we can again distinguish between maximal Newton polyhedra 
that have a direct product structure of the type 3+2+2 and 3+3,
and therefore inherit the connectivity of two dimensional polyhedra,
and the rest of them which are of the types 4+2, $\cont{\hbox{3+3}}$+2,
2+2+2+2, 5, $\cont{\hbox{4+3}}$, 
$\cont{\cont{\hbox{4+4}}}$ and $\cont{\hbox{3+3}\!\!}\!\cont{\hbox{~+3}}$.
These last classes are too large to be dealt with manually, so
we tackled the problem by computer.

The computer methods we are using are similar to the ones used in 
\cite{\rACJM}.
For the sake of completeness we will repeat part of the argument here and 
point out the differences. 
One possible approach is to try to identify all 
reflexive subpolyhedra (RSP) of each reflexive polyhedron (RP) that we try 
to connect.
\del
Finding one RSP contained in all RP's would solve the problem. 
\enddel
Since many of 
the RP's have over 200 points this decomposition is computationally 
prohibitive.

Another approach is to limit the search to a certain subset of RSP and see if 
there is one contained in all the RP's. 
This method also fails since there is no such magical RSP. 
To see this consider the minimal reflexive simplices 
that contain only 6 points. 
Since they cannot contain anything else they have 
to play the role of the magical RSP. 
There are nevertheless at least 3 minimal simplices in 4D that are 
inequivalent under the action of $GL(4,\IZ)$.
Of course it is not surprising that there is no single RSP
contained in every RP for $n=4$.
As we saw before, even in the much simpler case of $n=2$ we needed the
three RSP $\D_1^*$,  $\D_3^*$ and  $\D_4^*$ for showing connectedness.

For our proof we settled on the following strategy: 
identify as many 5-vertex irreducible RSP's as possible.
A 5-vertex irreducible polyhedron is a simplex that does not contain 
any other  reflexive simplex.
A set of 41 5-vertex irreducible simplices was generated in
\cite{\rACJM}, ranging in size from 6 to 26 points.
The list of 41 objects carries most of the burden of the connectedness
proof: the great majority of MNP contain at least one member of the list
and the 41 5-vertex irreducible simplices have already been proven to be 
connected \cite{\rACJM}. 
Instead of looking at 6 or 7-vertex irreducible objects as 
in \cite{\rACJM}
we dealt with the remaining polyhedra in the following two ways.

First note that $\Delta_1 \subset \Delta_2$ 
implies $\D_2^* \subset \D_1^*$. 
Suppose we want to show that $\Delta$ is connected to one of the 
41 5-vertex irreducible simplices, despite the fact that it doesn't contain 
any of them. The procedure is as follows.
We form the dual of $\Delta$, $\D^*$, and search for a 5-vertex irreducible 
simplex. 
If it contains one, 
say $\Delta^5_i$, it follows that $${(\D^5_i)}^* \supset \Delta.\eqlabel{@}$$ 
Finally, we need to search ${(\D^5_i)}^*$ for a 5-vertex irreducible simplex. 
If we find one, then ${(\D^5_i)}^*$ is connected to all of the others, and by 
\eqref{@}, so is $\Delta$. 
Applying this procedure to the remaining 305  MNP 
(235 of type 5, 68 of type $\cont{\cont{\hbox{4+4}}}$ and 2 of the type 3+3)
we succeeded in connecting all but two of them.

These last two, both of type 5, were treated by literally chopping them into 
smaller pieces. We did this by searching for hyperplanes that slice the 
polyhedra into two parts, and then checking the one that contains the 
interior point for reflexivity. In this way, one of the two troublesome 
polyhedra was reduced to a previously connected polyhedron (by the method 
described in the previous paragraph). The other was reduced to a polyhedron
that while not previously encountered, could be connected by the same method.

\section{co}{Concluding remarks}

The moduli space of  \cys\ is naturally divided into parts $\cal M$ 
where each manifold in the corresponding family can be obtained from any 
other manifold in the same family by smooth variations of the complex and 
K\"ahler structures.
\del
	The ultimate goal of all studies on connectedness is to show that all 
	of these families are connected in the sense that there are regions 
	of overlap between these a priori distinct moduli spaces.
\enddel
\new 
{}From the mathematical point of view the issue is to show that all
of these families are connected if boundary points of these distinct moduli
spaces, which correspond to singular varieties, are included.
Eventually it has to be checked that the physics of the involved singular
transitions is smooth for the various string theories that can be compactified
on the manifolds under consideration.
\endnew
\del
The first step of studying connectedness is to show that all
of these families are connected if boundary points of these distinct moduli 
spaces, which correspond to singular varieties, are included.
Eventually it has to be checked that the physics of the involved singular 
transitions is smooth for the various string theories that can be compactified
on the manifolds under consideration.
\enddel

The families we were studying are the ones that refer to \cy\ hypersurfaces in
toric varieties.
In this case each $\cal M$ contains one or more subspaces of the type 
${\cal M}_{(\D, \D^*)}$.
\del
	We have established that most of these moduli spaces are 
	mathematically	connected. 
\enddel
\new
We have established that all maximal polytopes defined by combined weight
systems with respect to the canonical (maximal) lattices, and therefore all
their subpolytopes, are connected. 
What still needs to be done in order to establish (mathematical) connectedness
of all toric \cys\ is to check connectedness of the maximal 
spanning polytopes that live on sublattices. We expect the number of new
polytopes that arise in this way to be rather small, but it requires a
considerable effort to construct all of them and we leave this task for future 
work.
\endnew

A particular moduli space $\cal M$ 	may 	
contain several "toric" 
submoduli spaces ${\cal M}_{(\D_i, \D^*_i)}$  
which will  					
partly span different dimensions of $\cal M$.%
\Footnote{ Even though the correct dimensions of 
	the homology groups can be calculated from the combinatorial data, 
	the lattice points
	in $(\D, \D^*)$ give us control on only part of the moduli space.}
Then, there will appear to be a  
transition (not necessarily extremal) where in fact we have just
changed from one toric description to another.
This is always the case for one and two dimensional Calabi--Yau 
spaces (tori and K3 surfaces, respectively),  but it may also happen
for threefolds. 
\del
We have established that most of the moduli space of \cys\ is mathematically
connected. 
It is naturally divided in subregions corresponding to families of 
\cys\ ${\cal M}_{(\D, \D^*)}$. 
Since the parameters of the
toric description that we use in our investigation span only part of 
the moduli space of \cy\ hypersurfaces, in some cases pairs of different 
polyhedra
define isomorphic families of such hypersurfaces even though the dimension of 
the 
part of the moduli space spanned by the toric parameters differs.
The submoduli spaces of topologically different families touch at points 
that represent singular varieties. 
This is not always the case when the transition 
takes place between different toric descriptions of the same family.
\enddel

Assuming there is full equivalence between 
$IIA[{\cal M}_{(\D, \D^*)}]$ and $IIB[{\cal M}_{(\D^*, \D)}]$ string theories
\cite{{{\rStro}, {\rAM}, {\rMori}}}  we will in the
following consider what happens when the connecting point is approached
by blow-downs in the ambient space.
Going from a family described by a polyhedron $\D_1^*$ to a family
described by a subpolyhedron $\D_2^* \subset \D_1^*$, we have to blow down 
one or
more divisors in the ambient space ${\cal V}_{\D_1^*}$.
Depending on the geometries of $\D_1^*$ and $\D_2^*$, 
such a divisor may be blown down to a surface, a curve, or a point.
Comparing this with general statements about boundaries of K\"ahler
moduli spaces for \cy\ threefolds \cite{{{\rWil}, {\rMV}}}, we see 
that the blow--down of a single T--divisor in the ambient space can 
have the following effects on the threefold:

\del
The blow down of a T-divisor in the 
embedding space may induce the 
following transformations in the \cy\ variety
\cite{{{\rWil}, {\rMV}}}: \hfil\break\ 
\enddel
\noindent
i) the family of hypersurfaces is unaffected.\hfil\break
\noindent
ii) a 2-parameter family of 2-cycles shrinks to a curve of 
singularities.\hfil\break
\noindent
iii) a 4-cycle shrinks to a point.

Let us start with discussing case (i). 
There are two essentially different ways in which a divisor in
the ambient space may be blown down without affecting the hypersurface.
One case is that the divisor does not intersect the hypersurface anyway.
This happens when the divisor corresponds to a lattice point in the 
interior of a facet of $\D_1^*$.
Since we are interested only in transitions induced by changes in the 
shapes of 
the polyhedra, this case will never characterize by itself a \hbox{3-fold} 
transition.
The other possibility is that a divisor is blown down to a surface that
is precisely the intersection of the original divisor with the \cy\ 
hypersurface.
This is a higher dimensional analog of what happens in our example of 
section 5.
Such a case was discussed in \cite{\rCDK}.

In the other two cases divisors of the \cy\ hypersurfaces are blown down.
As the canonical class will be affected in the process,
we cannot go from one smooth \cy\ manifold to another by this procedure.%
\Footnote{We thank S. Katz for clarifying discussions on this issue.}
When the transition involves a singular variety, this is reached 
from several different subregions by either moving to the boundary of 
the complex structure  or the \K\ class submoduli space. 
In these cases we deal with extremal transitions. 
Both  singularities described by cases (ii) and (iii) 
occur at finite distance in the moduli space \cite{\rHaya}.
In case (ii) we expect to experience an enhanced nonabelian gauge symmetry
\cite{{{\rAsp}, {\rAs}, {\rBSV}, {\rKM}, {\rKMP}, {\rBKKM}}} if and when 
the singularities have appropriate
properties while in case (iii) a generalization of the Argyres-Douglas
phenomenon is possible with the appearance of dyonic massless hypermultiplets
\cite{{{\rAD}, {\rCGGK}}}. 

In the majority of transitions more than one T-divisor has to be blown down
to realize the transition.
The superposition of the effects of individual blow-downs may lead in different
situations to an interplay between
cases (ii) and (iii). 
Moreover the curves of singularities might either develop 
their own singularities or may be shrunk to points by ulterior blow-downs,
or entirely new and unexpected phenomena might occur.

\del
Case (iii) may occur, as far as we know, when the transition takes place 
between
different toric representations of the same \cy\ family.
Even though we have not identified specific occurrences of transitions 
``between the same family" in our connectivity proof, 
this may happen. 
In such an instance the transition would take place between two different 
toric 
representations of the same family of \cy\ hypersurfaces that differ by the 
number
of polynomial (toric) deformations divisors) of the complex structure 
(in the ambient space). 

Since the transition implies a change in shape of the polyhedron we have to
blow-down divisors that do not belong to the interior of codimension one faces.
Since these will induce blow-downs of the \cy\ hypersurface the canonical
class will usually be affected in the process. 
By this procedure we cannot go from 
one smooth \cys\ to another unless the blow-down is an isomorphism 
\Footnote{We thank S. Katz for clarifying discussions on this issue}. 
Such cases where a divisor in the \cym\ is blown-down to a divisor in an 
isomorphic \cym\ have been discussed in \cite{\rCDK}.
\enddel
We end by pointing out that the connectedness of the moduli space of 3-folds
may have higher dimensional implications. It was  suggested  in \cite{\rBLS}
that by a  degeneration of the fibers, \cy\ 4-folds that are fibrations
with fiber a \cy\ 3-fold may be proven connected. This result is certainly 
true for the direct product spaces of the type $CY_4=CY_3 \times P^1$ 
where extremal
transitions between fibers are not constrained due to the double pyramidal
shape of the polyhedron.
\del
	It is yet to be proven that $CY_3$ fibrations in general inherit 
	the connectivity properties of their fibers since we don't know yet if 
	extremal transitions compatible with the fibration structure can be 
	found in all instances.
\enddel
\vskip10pt

{\it Acknowledgements:} 
We would like to thank P. Candelas and S. Katz for useful discussions. 
A.C.A. was supported in part by the Robert A. Welch Foundation and the NSF 
under grant PHY-9511632.
M.K.  was supported in part by the Austrian Research Funds FWF 
	grant P10641-PHY.
M.M. was supported by the United  States National Science Foundation
under grants PHY-9404057 and PHY-9457916.
H.S. is supported by the Austrian ``Fonds zur
F\"orderung der wissenschaftlichen Forschung'' with a Schr\"odinger
fellowship, number J012328-PHY
and by NSF grant PHY-9511632, the Robert A. Welch Foundation. 
\newpage
\font\bigtenrm=cmr10 scaled\magstep1
\def\skip{\hskip6pt\relax}
\baselineskip=4pt plus 2pt minus 2pt
\lineskip=4pt
\centerline{{\bigtenrm Table 1:} The 426 new spectra obtained from multiple 
		weight systems}
\def\spec#1#2#3{&  &#3  &&#1   &&#2  &\cr \noalign{\hrule}}
\def\begintable#1#2#3#4\endtable{%
	\line{\lineskip=10pt
	\vtop{
        \halign{\strut##&\vrule##&
	\hfil\skip\sixrm##\skip&##\vrule&
	\hfil\skip\sixrm##\skip&##\vrule&
	\hfil\skip\sixrm##\skip&##\vrule\cr
	\noalign{\hrule}
	& &\hbox{$\chi$}
    	&&$h_{11}$\hfil
      	&&$h_{21}$\hfil&\cr
	\noalign{\hrule\vskip3pt\hrule}#1
	\noalign {\hrule}}} 
        \hfill
	\vtop{
        \halign{\strut##&\vrule##&
	\hfil\skip\sixrm##\skip&##\vrule&
	\hfil\skip\sixrm##\skip&##\vrule&
	\hfil\skip\sixrm##\skip&##\vrule\cr
	\noalign{\hrule}
	& &\hbox{$\chi$}
    	&&$h_{11}$\hfil
      	&&$h_{21}$\hfil&\cr
	\noalign{\hrule\vskip3pt\hrule}
   #2
	\noalign {\hrule}}}
        \hfill	
        \vtop{
        \halign{\strut##&\vrule##&
	\hfil\skip\sixrm##\skip&##\vrule&
	\hfil\skip\sixrm##\skip&##\vrule&
	\hfil\skip\sixrm##\skip&##\vrule\cr
	\noalign{\hrule}
	& &\hbox{$\chi$}
    	&&$h_{11}$\hfil
      	&&$h_{21}$\hfil&\cr
	\noalign{\hrule\vskip3pt\hrule}
   #3
	\noalign {\hrule}}}
        \hfill	
        \vtop{
        \halign{\strut##&\vrule##&
	\hfil\skip\sixrm##\skip&##\vrule&
	\hfil\skip\sixrm##\skip&##\vrule&
	\hfil\skip\sixrm##\skip&##\vrule\cr
	\noalign{\hrule}
	& &\hbox{$\chi$}
    	&&$h_{11}$\hfil
      	&&$h_{21}$\hfil&\cr
	\noalign{\hrule\vskip3pt\hrule}
   #4
	\noalign {\hrule}}}

}
\vfill
}
        \hfill

\begintable{%
\spec{5}{197}{-384}\spec{6}{174}{-336}\spec{7}{175}{-336}\spec{6}{156}{-300}
\spec{8}{158}{-300}\spec{8}{154}{-292}\spec{6}{150}{-288}\spec{10}{152}{-284}
\spec{4}{142}{-276}\spec{7}{139}{-264}\spec{10}{142}{-264}\spec{11}{140}{-258}
\spec{5}{133}{-256}\spec{9}{135}{-252}\spec{9}{133}{-248}\spec{8}{130}{-244}
\spec{12}{134}{-244}\spec{5}{125}{-240}\spec{8}{128}{-240}\spec{12}{129}{-234}
\spec{4}{118}{-228}\spec{3}{115}{-224}\spec{4}{116}{-224}\spec{6}{118}{-224}
\spec{13}{125}{-224}\spec{5}{116}{-222}\spec{11}{121}{-220}\spec{3}{111}{-216}
\spec{7}{115}{-216}\spec{16}{124}{-216}\spec{11}{116}{-210}\spec{3}{107}{-208}
\spec{5}{109}{-208}\spec{10}{112}{-204}\spec{5}{104}{-198}\spec{12}{110}{-196}
\spec{4}{100}{-192}\spec{5}{99}{-188}\spec{9}{103}{-188}\spec{13}{106}{-186}
}{
\spec{16}{109}{-186}\spec{3}{95}{-184}\spec{8}{100}{-184}\spec{13}{103}{-180}
\spec{5}{93}{-176}\spec{12}{100}{-176}\spec{21}{109}{-176}\spec{13}{100}{-174}
\spec{15}{102}{-174}\spec{3}{89}{-172}\spec{17}{103}{-172}\spec{4}{88}{-168}
\spec{10}{93}{-166}\spec{5}{87}{-164}\spec{12}{94}{-164}\spec{2}{83}{-162}
\spec{6}{87}{-162}\spec{9}{90}{-162}\spec{12}{93}{-162}\spec{17}{98}{-162}
\spec{3}{83}{-160}\spec{10}{90}{-160}\spec{5}{84}{-158}\spec{8}{87}{-158}
\spec{3}{81}{-156}\spec{4}{82}{-156}\spec{13}{91}{-156}\spec{5}{81}{-152}
\spec{8}{84}{-152}\spec{5}{80}{-150}\spec{8}{83}{-150}\spec{16}{91}{-150}
\spec{20}{95}{-150}\spec{12}{86}{-148}\spec{14}{88}{-148}\spec{19}{93}{-148}
\spec{26}{99}{-146}\spec{4}{76}{-144}\spec{5}{76}{-142}\spec{7}{78}{-142}
}{
\spec{10}{81}{-142}\spec{5}{75}{-140}\spec{8}{78}{-140}\spec{17}{87}{-140}
\spec{11}{79}{-136}\spec{18}{86}{-136}\spec{8}{75}{-134}\spec{8}{74}{-132}
\spec{9}{75}{-132}\spec{13}{79}{-132}\spec{4}{68}{-128}\spec{5}{69}{-128}
\spec{9}{73}{-128}\spec{7}{70}{-126}\spec{9}{72}{-126}\spec{16}{79}{-126}
\spec{17}{80}{-126}\spec{21}{84}{-126}\spec{22}{85}{-126}\spec{8}{70}{-124}
\spec{10}{72}{-124}\spec{12}{74}{-124}\spec{11}{72}{-122}\spec{12}{73}{-122}
\spec{3}{63}{-120}\spec{7}{67}{-120}\spec{5}{64}{-118}\spec{6}{65}{-118}
\spec{7}{66}{-118}\spec{10}{69}{-118}\spec{11}{70}{-118}\spec{12}{71}{-118}
\spec{8}{66}{-116}\spec{14}{72}{-116}\spec{16}{74}{-116}\spec{13}{70}{-114}
\spec{17}{74}{-114}\spec{19}{76}{-114}\spec{21}{78}{-114}\spec{5}{61}{-112}
}{
\spec{8}{64}{-112}\spec{14}{70}{-112}\spec{8}{63}{-110}\spec{15}{70}{-110}
\spec{5}{59}{-108}\spec{12}{65}{-106}\spec{13}{66}{-106}\spec{14}{67}{-106}
\spec{15}{68}{-106}\spec{18}{71}{-106}\spec{28}{81}{-106}\spec{6}{58}{-104}
\spec{9}{61}{-104}\spec{12}{64}{-104}\spec{15}{67}{-104}\spec{9}{60}{-102}
\spec{14}{65}{-102}\spec{18}{69}{-102}\spec{20}{71}{-102}\spec{42}{93}{-102}
\spec{6}{55}{-98}\spec{11}{60}{-98}\spec{13}{62}{-98}\spec{4}{52}{-96}
\spec{7}{54}{-94}\spec{10}{57}{-94}\spec{12}{59}{-94}\spec{15}{62}{-94}
\spec{10}{56}{-92}\spec{14}{60}{-92}\spec{15}{61}{-92}\spec{19}{65}{-92}
\spec{10}{55}{-90}\spec{19}{64}{-90}\spec{20}{65}{-90}\spec{23}{68}{-90}
\spec{6}{50}{-88}\spec{7}{51}{-88}\spec{8}{52}{-88}\spec{15}{59}{-88}
}\endtable
\vfill\eject
\begintable{%
\spec{14}{57}{-86}\spec{16}{59}{-86}\spec{19}{62}{-86}\spec{7}{49}{-84}
\spec{12}{53}{-82}\spec{15}{56}{-82}\spec{7}{47}{-80}\spec{12}{52}{-80}
\spec{13}{53}{-80}\spec{6}{45}{-78}\spec{7}{46}{-78}\spec{8}{47}{-78}
\spec{9}{48}{-78}\spec{10}{49}{-78}\spec{12}{51}{-78}\spec{13}{52}{-78}
\spec{15}{54}{-78}\spec{16}{55}{-78}\spec{19}{58}{-78}\spec{21}{60}{-78}
\spec{25}{64}{-78}\spec{8}{46}{-76}\spec{14}{52}{-76}\spec{15}{53}{-76}
\spec{24}{62}{-76}\spec{27}{65}{-76}\spec{9}{46}{-74}\spec{11}{48}{-74}
\spec{12}{49}{-74}\spec{14}{51}{-74}\spec{17}{54}{-74}\spec{21}{58}{-74}
\spec{13}{48}{-70}\spec{15}{50}{-70}\spec{18}{53}{-70}\spec{20}{55}{-70}
\spec{6}{40}{-68}\spec{8}{42}{-68}\spec{11}{45}{-68}\spec{13}{47}{-68}
}{
\spec{14}{48}{-68}\spec{15}{49}{-68}\spec{17}{51}{-68}\spec{30}{64}{-68}
\spec{8}{41}{-66}\spec{12}{45}{-66}\spec{13}{46}{-66}\spec{15}{48}{-66}
\spec{19}{52}{-66}\spec{22}{55}{-66}\spec{30}{63}{-66}\spec{43}{76}{-66}
\spec{12}{44}{-64}\spec{18}{50}{-64}\spec{11}{42}{-62}\spec{12}{43}{-62}
\spec{14}{45}{-62}\spec{15}{46}{-62}\spec{16}{47}{-62}\spec{17}{48}{-62}
\spec{8}{38}{-60}\spec{12}{41}{-58}\spec{14}{43}{-58}\spec{16}{45}{-58}
\spec{18}{47}{-58}\spec{21}{50}{-58}\spec{22}{51}{-58}\spec{24}{53}{-58}
\spec{26}{55}{-58}\spec{10}{38}{-56}\spec{15}{43}{-56}\spec{13}{40}{-54}
\spec{14}{41}{-54}\spec{16}{43}{-54}\spec{23}{50}{-54}\spec{10}{36}{-52}
\spec{13}{38}{-50}\spec{18}{43}{-50}\spec{20}{45}{-50}\spec{28}{53}{-50}
}{
\spec{14}{37}{-46}\spec{21}{44}{-46}\spec{22}{45}{-46}\spec{31}{54}{-46}
\spec{13}{35}{-44}\spec{15}{37}{-44}\spec{19}{41}{-44}\spec{13}{34}{-42}
\spec{18}{39}{-42}\spec{19}{40}{-42}\spec{21}{42}{-42}\spec{24}{45}{-42}
\spec{30}{51}{-42}\spec{45}{66}{-42}\spec{12}{32}{-40}\spec{15}{34}{-38}
\spec{17}{36}{-38}\spec{19}{38}{-38}\spec{20}{39}{-38}\spec{22}{41}{-38}
\spec{23}{42}{-38}\spec{25}{44}{-38}\spec{30}{49}{-38}\spec{31}{50}{-38}
\spec{35}{54}{-38}\spec{37}{56}{-38}\spec{13}{31}{-36}\spec{14}{31}{-34}
\spec{15}{32}{-34}\spec{17}{34}{-34}\spec{19}{36}{-34}\spec{20}{37}{-34}
\spec{21}{38}{-34}\spec{26}{43}{-34}\spec{32}{49}{-34}\spec{14}{30}{-32}
\spec{18}{33}{-30}\spec{20}{35}{-30}\spec{22}{37}{-30}\spec{25}{40}{-30}
}{
\spec{30}{45}{-30}\spec{44}{59}{-30}\spec{18}{32}{-28}\spec{27}{41}{-28}
\spec{16}{29}{-26}\spec{17}{30}{-26}\spec{18}{31}{-26}\spec{24}{37}{-26}
\spec{28}{41}{-26}\spec{29}{42}{-26}\spec{32}{45}{-26}\spec{16}{27}{-22}
\spec{17}{28}{-22}\spec{19}{30}{-22}\spec{25}{36}{-22}\spec{26}{37}{-22}
\spec{30}{41}{-22}\spec{32}{43}{-22}\spec{16}{25}{-18}\spec{17}{26}{-18}
\spec{18}{27}{-18}\spec{30}{39}{-18}\spec{46}{55}{-18}\spec{15}{23}{-16}
\spec{16}{24}{-16}\spec{19}{26}{-14}\spec{25}{32}{-14}\spec{26}{33}{-14}
\spec{28}{35}{-14}\spec{33}{40}{-14}\spec{42}{49}{-14}\spec{16}{21}{-10}
\spec{18}{23}{-10}\spec{21}{26}{-10}\spec{22}{27}{-10}\spec{23}{28}{-10}
\spec{26}{31}{-10}\spec{27}{32}{-10}\spec{28}{33}{-10}\spec{30}{35}{-10}
}\endtable

\vfill\eject
\begintable{%
\spec{33}{38}{-10}\spec{35}{40}{-10}\spec{38}{43}{-10}\spec{15}{19}{-8}
\spec{16}{20}{-8}\spec{31}{34}{-6}\spec{34}{37}{-6}\spec{43}{45}{-4}
\spec{19}{20}{-2}\spec{20}{21}{-2}\spec{21}{22}{-2}\spec{22}{23}{-2}
\spec{25}{26}{-2}\spec{26}{27}{-2}\spec{30}{31}{-2}\spec{35}{36}{-2}
\spec{48}{49}{-2}\spec{30}{29}{2}\spec{39}{38}{2}\spec{40}{39}{2}
\spec{24}{21}{6}\spec{26}{23}{6}\spec{27}{24}{6}\spec{33}{30}{6}
\spec{41}{38}{6}\spec{47}{44}{6}
}{
\spec{22}{17}{10}\spec{24}{19}{10}
\spec{26}{21}{10}\spec{28}{23}{10}\spec{30}{25}{10}\spec{31}{26}{10}
\spec{32}{27}{10}\spec{35}{30}{10}\spec{37}{32}{10}\spec{41}{36}{10}
\spec{40}{33}{14}\spec{31}{22}{18}\spec{34}{25}{18}\spec{31}{21}{20}
\spec{24}{13}{22}\spec{26}{15}{22}\spec{28}{17}{22}\spec{29}{18}{22}
\spec{34}{23}{22}\spec{43}{32}{22}\spec{48}{37}{22}\spec{26}{13}{26}
\spec{29}{16}{26}\spec{32}{19}{26}\spec{34}{21}{26}\spec{40}{27}{26}
\spec{46}{33}{26}
}{
\spec{51}{38}{26}\spec{31}{16}{30}\spec{36}{21}{30}
\spec{58}{43}{30}\spec{30}{13}{34}\spec{35}{18}{34}\spec{36}{19}{34}
\spec{37}{20}{34}\spec{44}{27}{34}\spec{48}{31}{34}\spec{33}{14}{38}
\spec{34}{15}{38}\spec{36}{17}{38}\spec{39}{20}{38}\spec{43}{24}{38}
\spec{56}{37}{38}\spec{34}{11}{46}\spec{35}{12}{46}\spec{36}{13}{46}
\spec{41}{18}{46}\spec{42}{19}{46}\spec{43}{20}{46}\spec{51}{28}{46}
\spec{33}{8}{50}\spec{36}{11}{50}\spec{48}{23}{50}\spec{42}{13}{58}
}{
\spec{43}{14}{58}\spec{44}{15}{58}\spec{51}{22}{58}\spec{43}{12}{62}
\spec{48}{17}{62}\spec{58}{27}{62}\spec{52}{19}{66}\spec{45}{8}{74}
\spec{63}{24}{78}\spec{72}{33}{78}\spec{56}{13}{86}\spec{58}{15}{86}
\spec{66}{23}{86}\spec{53}{6}{94}\spec{59}{12}{94}\spec{68}{21}{94}
\spec{58}{9}{98}\spec{70}{21}{98}\spec{63}{8}{110}\spec{71}{10}{122}
\spec{78}{17}{122}\spec{68}{5}{126}\spec{72}{9}{126}\spec{75}{8}{134}
\spec{101}{27}{148}\spec{112}{32}{160}
}\endtable

\vfill\eject
\centerline{{\bigtenrm Table 2:} Results for multiple weight systems. 
	$\#sym$ refers to comparing mirror pairs of 	}
\centerline{~~~~~ ~~ ~~ ~~ 
	spectra. $\#new'$ and $\#sym'$ give the respective
	numbers of new spectra as com- }
\centerline{~~~~~ ~~~~~~
	pared to single weight systems $and$ abelian orbifolds of 
	transversal weights \cite{\rAAS}}
						\let\sixrm=\relax
\bigskip
\centerline{
\vtop{
\halign{\strut#&\vrule#&
\skip\sixrm#\skip\hfil&#\vrule&
\hfil\skip\sixrm#\skip&#\vrule&
\hfil\skip\sixrm#\skip&#\vrule&
\hfil\skip\sixrm#\skip&#\vrule&
\hfil\skip\sixrm#\skip&#\vrule&
\hfil\skip\sixrm#\skip&#\vrule&
\hfil\skip\sixrm#\skip&#\vrule\cr
\noalign{\hrule}
&
 &\hbox{\sixrm Weight systems}
  &&\hbox{\sixrm \# poly}
    &&\hbox{\sixrm \# spec}\hfil
      &&\hbox{\sixrm \# new}\hfil
        &&\hbox{\sixrm \# sym}\hfil
          &&\hbox{\sixrm \# new'}\hfil
            &&\hbox{\sixrm \# sym'}&\cr
\noalign{\hrule\vskip3pt\hrule}
& &$\cont{\cont{\hbox{4+4}}}$  &&6365  &&2078  &&381  &&101  &&332  &&96 &\cr
\noalign{\hrule}
& &$\cont{\hbox{4+3}}$    &&727   &&485    &&73    &&9    &&60   &&7 &\cr 
\noalign{\hrule}
& &4+2     &&58    &&56     &&8     &&0    &&6    &&0 &\cr
\noalign{\hrule}
& &$\cont{\hbox{3+3}\hskip -3pt}\hskip -2pt\cont{\hbox{~+3}}$   &&36    &&29 
	   &&5     &&0    &&4    &&0 &\cr
\noalign{\hrule}
& &$\cont{\hbox{3+3}}$+2  &&17    &&17     &&4     &&0    &&4    &&0 &\cr
\noalign{\hrule}
& &3+3     &&6     &&6      &&3     &&1    &&3    &&1 &\cr
\noalign{\hrule}
& &3+2+2   &&3     &&3      &&1     &&0    &&1    &&0 &\cr
\noalign{\hrule}
& &2+2+2+2 &&1     &&1      &&1     &&0    &&1    &&0 &\cr
\noalign{\hrule}
& &total   &&7213  &&2171   &&426   &&105  &&373  &&100 &\cr 
\noalign{\hrule}}}}

\newpage
\baselineskip=12pt plus 1pt minus 1pt
\immediate\closeout\referencewrite
\referenceopenfalse
\line{\bf\hfil References\hfil}\vskip.2truein
\input referenc.texauxil

\bye